\documentclass[aps,prapplied,amsmath,amssymb,twocolumn,superscriptaddress]{revtex4-2}

\usepackage{graphicx}
\usepackage{adjustbox}
\usepackage{fixltx2e}
\usepackage{makecell}
\usepackage{hyperref}
\usepackage{cleveref}
\usepackage{subcaption}
\graphicspath{ {./pictures/} }

\begin{document}


\title{CMOS-compatible processing and room-temperature characterization on wafer-level for scalable quantum computing}


\author{S. J. K. Lang*}
\email{simon.lang@emft.fraunhofer.de}
\affiliation{Fraunhofer Institut für Elektronische Mikrosysteme und Festkörpertechnologien EMFT, Munich, Germany}
\author{T. Mayer*}
\email{thomas.mayer@emft.fraunhofer.de}
\affiliation{Fraunhofer Institut für Elektronische Mikrosysteme und Festkörpertechnologien EMFT, Munich, Germany}
\author{J. Weber*}
\email{johannes.weber@emft.fraunhofer.de}
\affiliation{Fraunhofer Institut für Elektronische Mikrosysteme und Festkörpertechnologien EMFT, Munich, Germany}
\author{C. Dhieb}
\affiliation{Fraunhofer Institut für Elektronische Mikrosysteme und Festkörpertechnologien EMFT, Munich, Germany}
\author{I. Eisele}
\affiliation{Fraunhofer Institut für Elektronische Mikrosysteme und Festkörpertechnologien EMFT, Munich, Germany}
\affiliation{Center Integrated Sensor Systems (SENS), Universität der Bundeswehr München, Munich, Germany}
\author{W. Lerch}
\affiliation{Fraunhofer Institut für Elektronische Mikrosysteme und Festkörpertechnologien EMFT, Munich, Germany}
\author{Z. Luo}
\affiliation{Technical University of Munich, Munich, Germany}
\author{C. Morán Guizán}
\affiliation{Fraunhofer Institut für Elektronische Mikrosysteme und Festkörpertechnologien EMFT, Munich, Germany}
\author{E. Music}
\affiliation{Fraunhofer Institut für Elektronische Mikrosysteme und Festkörpertechnologien EMFT, Munich, Germany}
\author{L. Sturm-Rogon}
\affiliation{Fraunhofer Institut für Elektronische Mikrosysteme und Festkörpertechnologien EMFT, Munich, Germany}
\author{D. Zahn}
\affiliation{Fraunhofer Institut für Elektronische Mikrosysteme und Festkörpertechnologien EMFT, Munich, Germany}
\author{R.N. Pereira}
\affiliation{Fraunhofer Institut für Elektronische Mikrosysteme und Festkörpertechnologien EMFT, Munich, Germany}
\author{C. Kutter}
\affiliation{Fraunhofer Institut für Elektronische Mikrosysteme und Festkörpertechnologien EMFT, Munich, Germany}
\affiliation{Center Integrated Sensor Systems (SENS), Universität der Bundeswehr München, Munich, Germany}
\collaboration{*These authors contributed equally to this work and are listed in alphabetical order.}



\date{\today}

\begin{abstract}
We report on an industry-grade CMOS-compatible qubit fabrication approach using a CMOS pilot line, enabling a yield of functional devices reaching 92.8\,\%, with a resistance spread evaluated across the full wafer 200\,mm diameter of 12.4\,\% and relaxation times ($T_1$) approaching 80\,µs.  Furthermore, we conducted a comprehensive analysis of wafer-scale room temperature (RT) characteristics collected from multiple wafers and fabrication runs, focusing on RT measurements and their correlation to low temperature qubit parameters. From defined test structures, an across-wafer Josephson junction (JJ) area variation of 10.1\,\% and oxide barrier variation of 7.2\,\% was calculated. Additionally, from the room-temperature JJ characterization the qubit frequency can be derived on wafer-level applying the Ambegaokar-Baratoff model before low temperature measurements. This sets the stage for pre-cooldown wafer-level JJ evaluation and sorting. In particular, such early-on device characterization and validation are crucial for increasing the fabrication yield and qubit frequency targeting, which currently represent major scaling challenges. Furthermore, it enables the fabrication of large multichip quantum systems in the future. Our analysis highlight the great potential of CMOS-compatible industry-style fabrication of superconducting qubits for scalable quantum computing in a foundry pilot line cleanroom.
\end{abstract}


\maketitle
\section{Introduction} 
Scalable, CMOS-compatible processing on 200\,mm wafers is a critical step toward industrializing superconducting qubit-based quantum processing units \cite{mohseni_How_2025}. It enables the precise fabrication and complex 3D integration needed for advanced quantum architectures \cite{Mayer_3D_2025, Rosenberg_3D_2017}. By leveraging mature semiconductor manufacturing methods, this approach ensures reproducibility, scalability, and the transition from laboratory prototypes to large-scale quantum systems. The fabrication of CMOS-compatible superconducting all-aluminum junctions using a double dry-etch technique with industry-standard manufacturing tools has been successfully demonstrated \cite{VanDamme_Advanced_2024, Stehil_Coherent_2020, lang_aluminum_2023}. In addition, comprehensive wafer-scale characterization at RT has been reported \cite{Wan_Fabrication_2021, Lang_room_2025}. A recent study \cite{VanDamme_Advanced_2024} integrated both approaches, showcasing the benefits of industry-compatible processes for transmon-style qubit fabrication, achieving a high yield of 98.5\,\% and relaxation times ($T_1$) of up to 167\,µs. The study indicated that junction area and oxide barrier are essential parameters for the precise frequency targeting. As they vary across the wafer, they are worth monitoring for improved device characterisation. Additionally, like others \cite{pishchimova_improving_2022, Osman_Simplified_2021}, they used critical dimension scanning electron microscopy (CD-SEM) to evaluate their structuring process at RT. Subsequently, they analyzed selected qubits after cooldown in the cryostat for benchmarking. This low temperature characterization, which is often very time-consuming and performed at die-level after leaving the production line, has become an established method for qubit characterization \cite{siddiqi_engineering_2021}. Currently, there is significant research focused on developing time-efficient, wafer-scale characterization methods for qubits during fabrication as part of the process control monitoring (PCM). This approach relies on RT techniques, as traditional cryogenic methods are impractical for large-scale analysis. Simple resistance measurements on test structures, like shorts and test junctions, can be used to monitor the line width variation as well as the junction oxide barrier variation across the 200\,mm wafer. Additionally, a selection for functional qubit chips can be implemented to increase the yield for the subsequent process steps, as non-functional devices can be excluded early on with high probability. The primary aim of this wafer-scale characterization is to predict the cryogenic behavior of qubit devices, including the qubit frequency targeting, based on measurements conducted at RT. 
\\ \\
This present study demonstrates a comprehensive methodology for the fabrication and RT characterization of superconducting qubits on a 200 mm silicon wafer, emphasizing the importance of detailed wafer-level analysis for process monitoring. By integrating an array of test structures, including test junctions and shorts, our approach facilitates the precise evaluation of junction characteristics at RT and variation across the entire 200 mm wafer. For example, the analysis of shorts enabled an efficient evaluation of the metal structuring variability in terms of the junction area variation without using CD-SEM. Measuring the junctions resistance enabled an accurate selection of functional devices, which had a high yield across the full 200 mm wafer. In combination with the cryogenic analysis of those functional qubit chips, we could demonstrate validity of the Ambegaokar-Baratoff relation for all our samples over a large resistance range. This enables us to select qubit chips based on their resistance already at RT at wafer-level. Among others, this approach is especially important for further processing in 3D integrated architectures for increased yield. Our study highlights the effectiveness of the proposed characterization principles for CMOS-compatible superconducting qubits, which are crucial for achieving time-efficient process control and optimization for scaling of a QPU. 

\section{Methods}
\begin{figure*}[htbp]
\centering
\begin{minipage}{\textwidth}
\includegraphics[height=5cm]{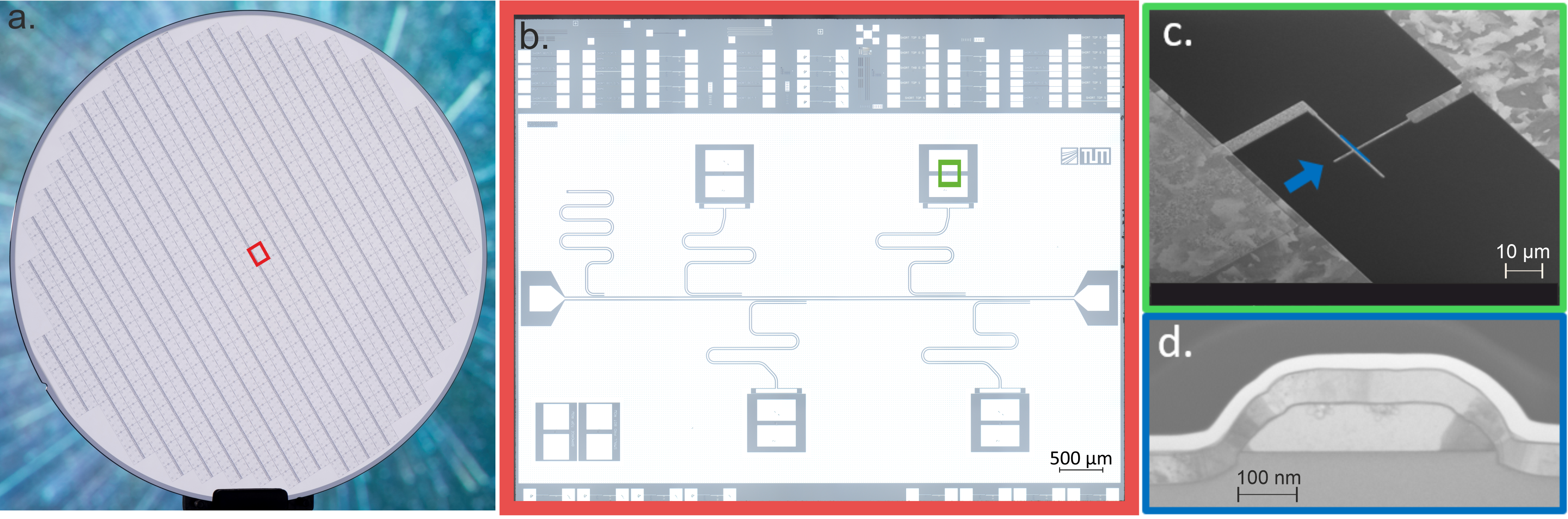}
\caption{a. 200 mm qubit wafer with 106 dies. b. Layout of a single die with qubits and test structures. c. SEM image of a qubit’s JJ. d. Electron microscopy image of the cross section of a JJ.}
\label{fig:SC}
\end{minipage}
\end{figure*}

In total 375 identical dies with a size of 10$\,$mm x 7$\,$mm containing three different sub-dies were fabricated on a single wafer (Fig. \ref{fig:SC}a). The relevant sub-die, shown in Fig. \ref{fig:SC}b, is the qubit chip with a single reference resonator and four fixed-frequency qubits with equal Josephson junction (JJ) area, all addressed through a common RF feedline. Additionally, a process control monitoring (PCM) array was placed in the wafer’s dicing street. It consist of 16 test junctions of sizes varying from 0.1225$\,$µ$\text{m}^2$ up to 1$\,$µ$\text{m}^2$ and eight shorts (i.e. simple metal lines with landing pads for the prober needles) with widths from 0.35$\,$µm to 1$\,$µm. The fixed-frequency qubits were used to analyze the cryogenic performance of the fabrication process by measuring the qubits transition frequencies ($f_{\text{qb}}$), relaxation times ($T_1$), and dephasing times ($T_2^*$). With the test junctions, RT resistance and IV measurements were performed to access junction’s resistances and breakthrough voltage without harming the qubits JJ. Additionally, the shorts were used to obtain information about the JJ electrode width distribution across the 200$\,$mm wafer, which is limited by the structuring process. Both test structures were used to maintain a time-efficient RT PCM during fabrication. Additionally, the sheet resistance of the bottom Al layer was measured using a standard four-point probe (4pp) setup to eliminate contact resistance. A standard 49-point sampling pattern across the 200\,mm wafer was used to assess uniformity.

\begin{figure*}[htbp]
    \centering
    \begin{minipage}{0.32\textwidth}
        \centering
        \raisebox{0cm}{\includegraphics[height=2.5cm]{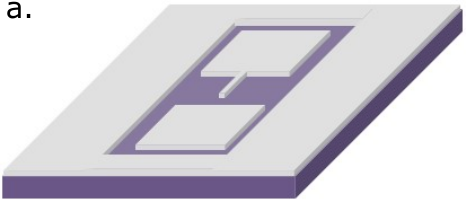}}
    \end{minipage}%
    \hspace{0.01\textwidth} 
    \begin{minipage}{0.32\textwidth}
        \centering
         \raisebox{1.7cm}{\includegraphics[height=4.3cm]{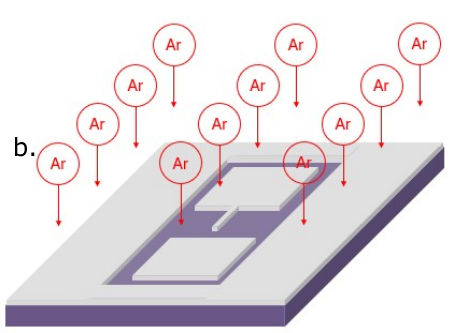}}
    \end{minipage}%
    \hspace{0.01\textwidth}
    \begin{minipage}{0.32\textwidth}
        \centering
        \includegraphics[height=2.5cm]{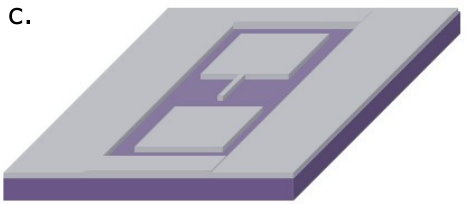}
    \end{minipage}
    
    \vspace{0cm} 

    \centering
    \begin{minipage}{0.32\textwidth}
        \centering
        \includegraphics[height=2.6cm]{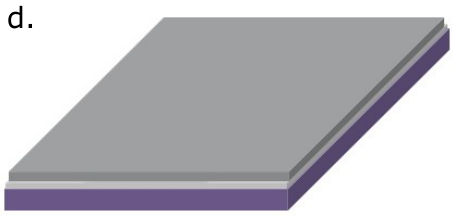}
    \end{minipage}%
    \hspace{0.04\textwidth}
    \begin{minipage}{0.32\textwidth}
        \centering
        \includegraphics[height=2.5cm]{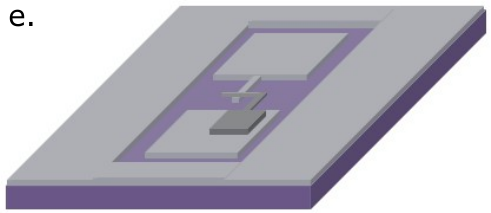}
    \end{minipage}
    \caption{Qubit fabrication schematic: a. structuring of BE based on Al (grey) on Si (purple), b. Ar ion milling of BE to remove native oxide, c. static oxidation of BE to form junction oxide, d. capping with top Al layer (dark grey), e. structuring of TE.}
    \label{fig:process}
\end{figure*}

The CMOS-compatible front-end fabrication was performed in the 200\,mm pilot line cleanroom at Fraunhofer EMFT, utilizing (100) p-type silicon wafers with a high resistivity of 3-5\,$\text{k}\Omega \text{cm}$ and a previously reported overlap fabrication process for the Al/AlOx/Al junctions \cite{Stehil_Coherent_2020, Wu_Overlap_2017}. The applicability of this process has been recently demonstrated for CMOS technology \cite{lang_aluminum_2023, verjauw_path_2022, VanDamme_Advanced_2024}. Before depositing the first 150\,nm thick Al layer using a sputter tool, an ex-situ liquid HF dip was performed for 3\,min followed by dry cleaning in an N2 atmosphere. Afterwards, the first (bottom) Al layer was patterned using an optical lithography process in an i-line stepper (365\,nm). With a dry etch using chlorine chemistry, the resist layout was transferred to the Al film. Subsequently, the resist was ashed with $\text{H}_2\text{O}$ and $\text{O}_2$ plasma, followed by a residual cleaning procedure using EKC, isopropanol and de-ionized (DI) water, completing the structuring of the JJ bottom electrode (BE) and all other required structures like capacitive pads, feedline, resonators and grounding layer (Fig. \ref{fig:process}a).

The JJ oxide was formed by performing the next three subsequent process steps in-situ. First, the native AlOx barrier on the BE was removed with Ar ion milling (Fig. \ref{fig:process}b), followed by static oxidation with $\text{O}_2$ (Fig. \ref{fig:process}c). Afterwards, the oxide was directly capped by the top Al layer (100\,nm thick) to avoid reoxidation (Fig. \ref{fig:process}d). The top Al layer was deposited under the same sputter conditions as the bottom layer (described above). To finalize the junction fabrication, the top electrode (TE) was patterned similarly to the BE (Fig. \ref{fig:process}e). The electrical connection between the TE of the junction and the qubit capacitor pad was provided by a large-area junction (Fig. \ref{fig:process}e and Fig. \ref{fig:SC}c) \cite{verjauw_path_2022}.
\\ \\
This second junction with a size of 2500$\,$µm$^2$ was fabricated simultaneously during the TE structuring process to minimize the number of process steps as no additional contact metal layer was needed. After fabrication, a PCM routine was conducted, involving 4-point resistance measurements on the shorts and 2-point current-voltage (IV) measurements on the test junctions, using a fully automated wafer probe station. Here, the voltage applied to the test junctions was step-wise increased by 10\,mV intervals up to the breakthrough of the JJs, while measuring the current. After completing this wafer-level characterization, the wafers were diced. Here, a protective resist was applied, which was removed afterward for each die individually using acetone, isopropanol, and DI water. 

\section{PCM with test structures}
Based on our process, we were able to derive methodologies for an RT SPC (statistical process control) for the CMOS-compatible fabrication of superconducting qubits, described in the following section. The usefulness of our approach is twofold: it provides valuable insights for process development and optimization and enables the identification of qubits that matches the design parameters already at RT, before cryo-measurements were carried out. In section 4, we demonstrate this by reproducibly determining the qubit frequencies at RT from the junction resistances using the Ambegaokar-Baratoff formula.

\subsection{Shorts Analysis}

\begin{figure}[h!]
\centering
    \begin{subfigure}[b]{0.95\linewidth}
        \centering
        \includegraphics[height=5.8cm]{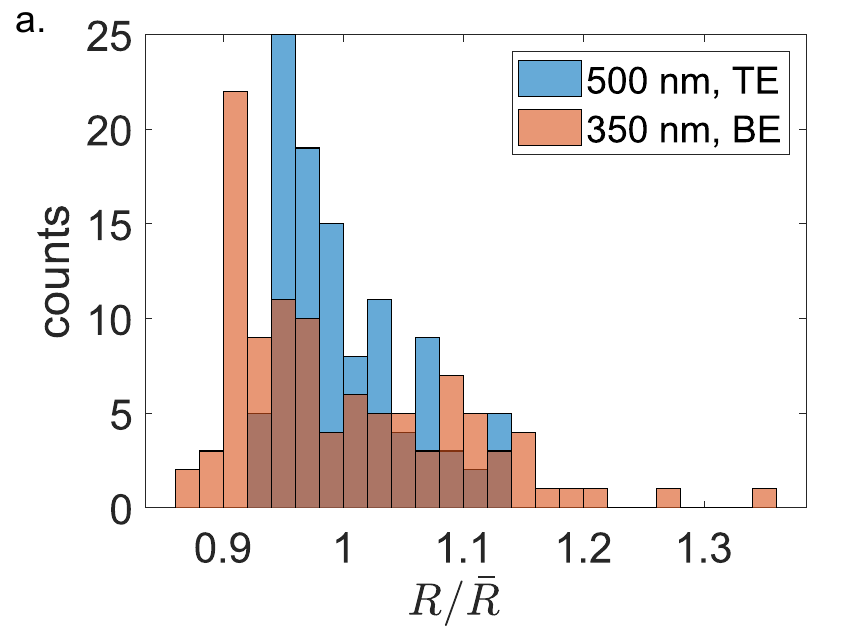}
        \label{fig:3sub1}
    \end{subfigure}
\vskip\baselineskip
    \begin{subfigure}[b]{0.95\linewidth}
        \centering
        \includegraphics[height=5.8cm]{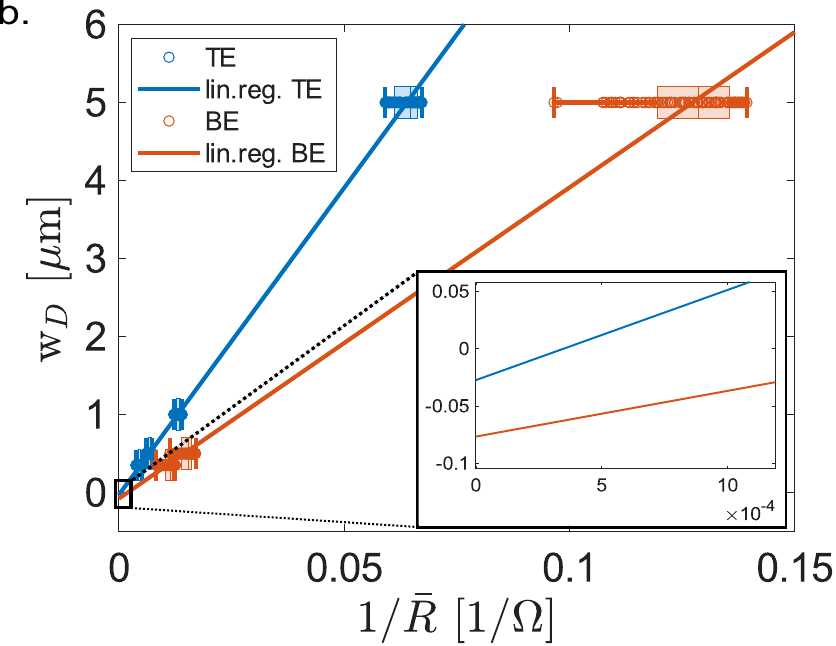}
        \label{fig:1sub2y}
    \end{subfigure}
\vskip\baselineskip
    \begin{subfigure}[b]{0.95\linewidth}
        \centering
        \includegraphics[height=6cm]{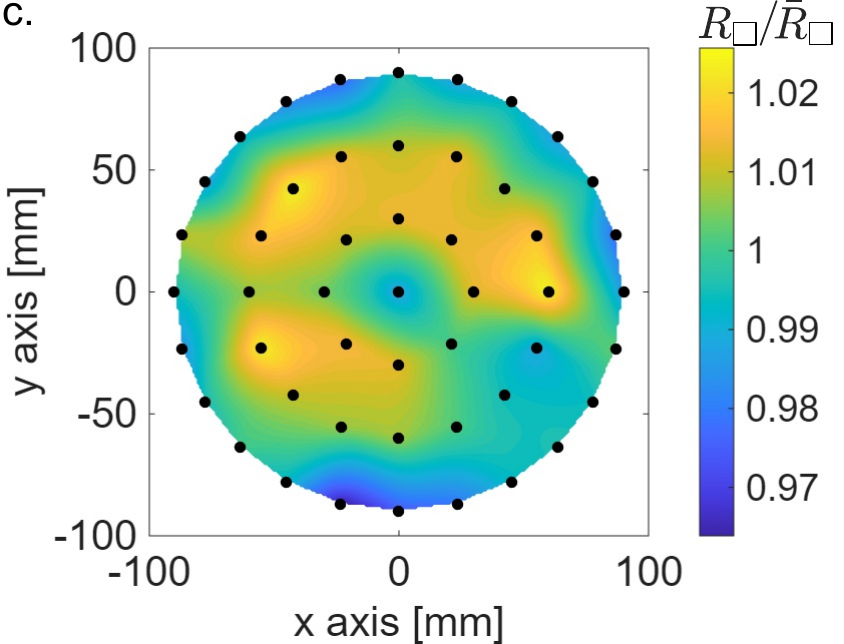}
        \label{fig:1sub2y}
    \end{subfigure}
\caption{a. Normalized resistance ($R/\bar{R}$) histograms obtained for BE shorts (width of 350\,nm and $RSD_\text{w}^{\text{BE}}$ = 8.8\,\%) and for TE shorts(width of 500\,nm and $RSD_\text{w}^{\text{TE}}$  = 5.3\,\%). b. design width $w_D$ vs. across-wafer inverse resistance mean 1/$\bar{R}$. With a linear fit, the offset $\Delta \text{w}$ between design and fabrication width can be extracted by the crossing at the y-axis (zoomed in). c. Normalized sheet resistance map ($R_\square/\bar{R}_\square$) from a 49-point 4pp measurement on a 150\,nm thick Al layer, showing a relative standard deviation of $RSD_{\text{R}_\square}$\,=\,1.4\,\%.}
\label{fig:short}
\end{figure}

To electrically examine the width variability of the metal structuring process, shorts fabricated simultaneously with the TE and BE layers of the JJs were analyzed. To compare the results from the shorts with data obtained for the qubit JJs (350\,nm $\times$ 500\,nm), shorts with identical sizes were used. Specifically, we analyzed the resistance $R$ of shorts with a width of 350\,nm, fabricated during the BE structuring, and shorts with a width of 500\,nm, fabricated during the TE structuring. The cross-wafer mean resistance $\bar{R}$ obtained for the BE shorts was 94.7\,$\Omega$, while for the TE shorts, it was 149.6\,$\Omega$. Fig. \ref{fig:short}a shows the distribution of the resistances, normalized to their respective mean values ($R/\bar{R}$), obtained for the two short sizes. From those distributions, the relative standard deviation (RSD) of the top and bottom short $RSD_\text{R}^{\text{TE}/\text{BE}}$ can be extracted. With the RSD of the sheet resistance $RSD_{\text{R}_\square}$, we can calculate the RSD of the widths of the top short $RSD_\text{w}^{\text{TE}}$ and bottom short $RSD_\text{w}^{\text{BE}}$ by
\begin{equation}
RSD_\text{w}^{\text{TE}/\text{BE}} = \sqrt{\left( RSD_\text{R}^{\text{TE}/\text{BE}}\right)^2  - \left(RSD_{\text{R}_\square}\right)^2 }
\end{equation}
In general, for all our Al depositions for the layer thicknesses of 150\,nm and 100\,nm we measured $RSD_{\text{R}_\square}$ of $\leq$\,1.6\,\%. As an example, Fig.~\ref{fig:short}c shows a 49-point standard 4pp measurement performed on a 200\,mm wafer to determine the sheet resistance of a 150\,nm thick Al film. Therefore, we assumed this value for the BE and TE deposition for the samples in this paper. For this, we obtain a value of $RSD_\text{w}^{\text{TE}}$\,=\,5.1\,\%, which is smaller than the BE width distribution of $RSD_\text{w}^{\text{BE}}$\,=\,8.7\,\%. Such increase in width variability for decreasing design width is frequently observed for metal structuring \cite{Osman_Simplified_2021}. Additionally, a $RSD_\text{A}$ = 10.1\,\% for the area variation is obtained from the relation
\begin{equation}
RSD_\text{A} = \sqrt{\left( RSD_\text{w}^{\text{TE}}\right)^2  + \left(RSD_\text{w}^{\text{BE}}\right)^2}
\end{equation}
The $RSD_\text{A}$, observed for our junctions, with an area of 0.175\,µm$^2$, is notably larger than compared to values ($RSD_\text{A} \approx 3.8\,\%$) reported sizes for JJs with areas (A $\approx$ 0.04\,µm$^2$) typically used in superconducting qubits \cite{VanDamme_Advanced_2024, Osman_Simplified_2021, pishchimova_improving_2022}. When using optical lithography, the design widths approach the limits of the i-line stepper (365\,nm) tool, which causes an increase in area variability. Using electron-beam lithography for junction patterning circumvents this limitation. We note that our time-efficient determination of electrode width variability, which applies only RT measurements of test structures, is much faster than the time-consuming optical inspection with critical dimension scanning electron microscope (CD-SEM) that is commonly used. 

 Furthermore, additional shorts with different widths can be used to determine the offset $\Delta \text{w}$ between the design width $\text{w}_\text{D}$ and actual width w. As the across-wafer resistance mean values $\bar{R}$ of the respective shorts is inverse proportional to w, we can define $\text{w}_\text{D} - \Delta \text{w} = \text{w} \sim 1 / \bar{R}$. This approach can be used for both metal layers separately. The offset of the structuring process to the design value can be derived from Fig. \ref{fig:short}b at the crossing of the fit with the y-axis and is $\Delta \text{w}_{\text{BE}}$  = - 76.8\,$\pm$\,18.5\,nm for the BE and $\Delta \text{w}_{\text{TE}}$  = - 27.5\,$\pm$\,7.2\,nm for the TE. These offsets result from approaching the feature-size limit of the optical lithography, which can be circumvented by electron-beam lithography as mentioned before. Thus, our PCM of shorts enable not only the determination of the area size variability across the wafer ($RSD_\text{A}$) in a time-efficient manner, but also provides a calculation of the offset from the design widths $\Delta \text{w}^{\text{TE}/\text{BE}}$ .  

\subsection{Junction analysis}

\begin{figure}[h!]
\centering
    \begin{subfigure}[b]{0.95\linewidth}
        \centering
        \includegraphics[height=6cm]{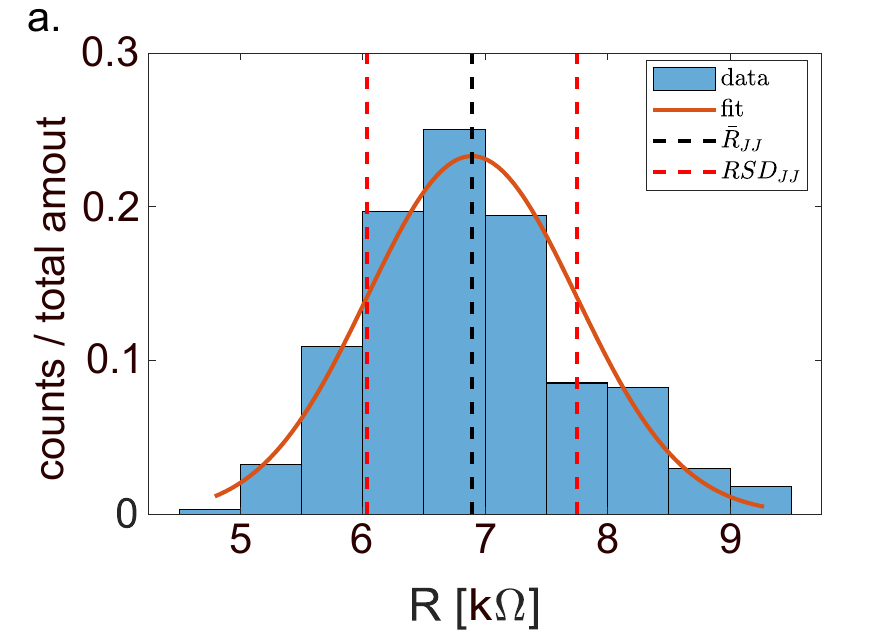}
        \label{fig:4sub1}
    \end{subfigure}
\vskip\baselineskip
    \begin{subfigure}[b]{0.95\linewidth}
        \centering
        \includegraphics[height=6cm]{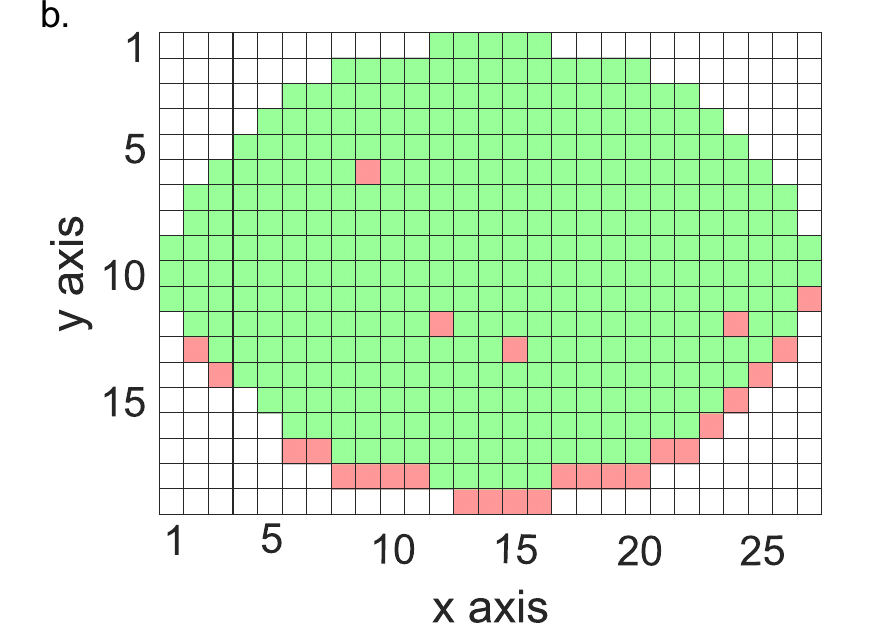}
        \label{fig:1sub2}
    \end{subfigure}
\caption{histogram (a.) and yield wafer map (b.) of $R_{\text{JJ}}$ with junction area of 0.175\,µm$^2$ across wafer with specification 100\,$\Omega < R_{\text{JJ}} < 50$\,k$\Omega$. }
\label{fig:JJs}
\end{figure}

Besides the shorts, we also fabricated test JJs as PCM test structures during the qubit fabrication. This enables an approximation of the qubits JJ’s resistance prior to cryogenic characterization. As our qubit design aimed for a qubit frequency of 4.42 GHz and the critical temperature of the Al JJ was fitted to be about 0.7\,K, according the Ambegaokar-Baratoff relation \cite{ambegaokar_tunneling_1963}, a RT junction resistance $R_{\text{JJ}}$ of about 7.3\,k$\Omega$ is targeted for the designated junction area. The results of the cryogenic measurements will be shown and analyzed in the next sections, after the discussion of the RT measurements. As a second stage of the PCM, the across-wafer RT resistance distribution of test junctions identical to the qubits JJ were measured and analyzed (Fig. \ref{fig:JJs}a). The test junctions are particularly useful because they can be placed on a chip in larger numbers without the metal pads and shielding needed for qubits. At the same time, the overlap area of the JJs can be varied.
\\ \\ 
In the context of future deployment of our qubits in quantum processing units, it is evident that sorting is essential to exclude unsuitable devices. Therefore, we applied a specification for our devices according to their resistance to exclude short and open circuits, which is defined as  100\,$\Omega < R_{\text{JJ}} < 50$\,k$\Omega$. In Fig. \ref{fig:JJs}b, a wafer map is shown, with the junctions fulfilling the requirements marked green (working junctions), and the failed ones depicted in red. The failed junctions tend to be distributed randomly across the wafer except for the bottom wafer edge. The yield $Y_R$, defined by the ratio of the working junctions and the total number of junctions, was recorded to be 92.8\,\%. 

\begin{figure}[h]
\centering
    \begin{subfigure}[b]{0.95\linewidth}
        \centering
        \includegraphics[height=5.9cm]{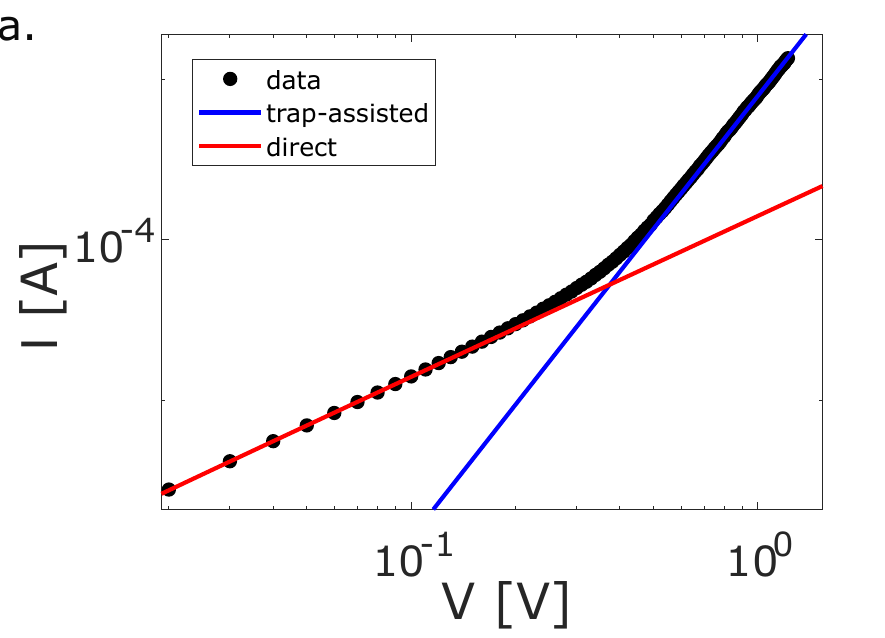}
        \label{fig:1sub1}
    \end{subfigure}
\vskip\baselineskip
    \begin{subfigure}[b]{0.95\linewidth}
        \centering
        \includegraphics[height=6cm]{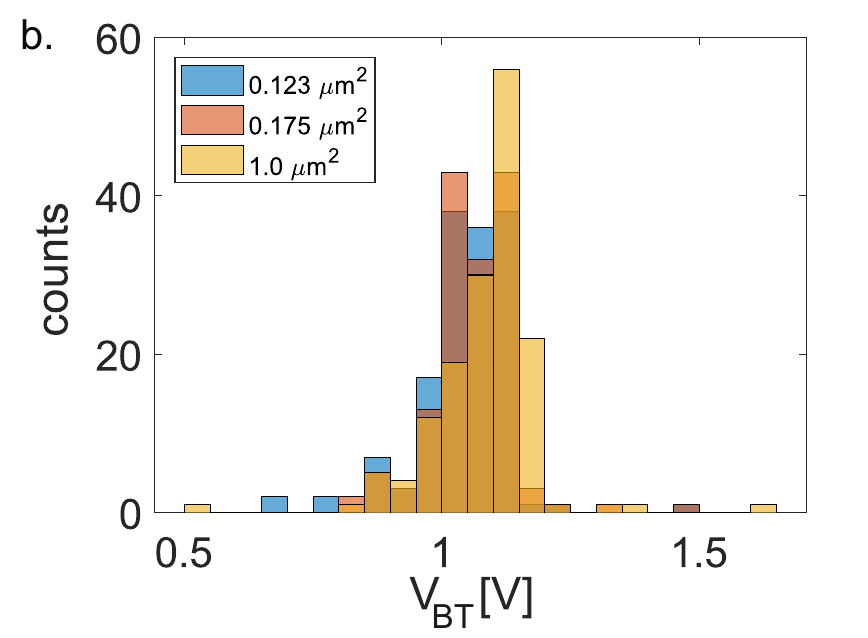}
        \label{fig:1sub2x}
    \end{subfigure}
\caption{a. Typical IV-curve recorded for the test junctions of size 0.175\,µm$^2$ with direct (red) tunneling and trap assisted (blue) tunneling with $m$ = 2.75 before the breakthrough at 1.2\,V. A value of $R_{\text{JJ}}$ = 7.1\,k$\Omega$ was measured in the direct tunneling regime. b. Histogram of $V_{\text{BT}}$ for different junction sizes.  }
\label{fig:IV}
\end{figure}

These junctions were further selected with the 1.5*IQR rule (interquartile range), a common criteria for proper data analysis. On this account, the histogram (Fig. \ref{fig:JJs}a) of the across-wafer RT resistance distribution can be fitted by a Gaussian distribution with a deviation $RSD_{\text{JJ}}$ of 12.4\,\%. Notably, the mean value of the distribution with $\bar{R_{\text{JJ}}}$ = 6.89\,k$\Omega$ fits the target value of 7.3\,k$\Omega$ within 5.6\,\%. In combination with the $RSD_\text{A}$, obtained from the shorts (see Sec. 3.1), the joint RSD for the junction oxides resistivity $\rho$ and thickness $t$, can be calculated with
\begin{equation}
RSD_{\text{RA}} = \sqrt{ RSD_{\text{JJ}}^2  - RSD_\text{A}^2}
\end{equation}
$RSD_{\text{RA}}$ = 7.2\,\% is smaller than the area spread of $RSD_\text{A}$ = 10.1\,\%, indicating that the contribution of the spread in oxide thickness and barrier height is not dominant, although it remains larger than in related works \cite{VanDamme_Advanced_2024}. By optimizing junction oxidation and Ar milling parameters, these variations can be further reduced. However, the miminum achievable resistance variation across the wafer needs to be further investigated. To further analyzing our junctions in terms of their tunnel barrier characteristics extended current voltage measurements were carried out. An example is shown in Fig. \ref{fig:IV}a. In general, the current $I$ through the junction can be described with a simple power-law dependence \cite{Chiang_electronic_2018}
\begin{equation}
I \sim V^m
\end{equation}
For low voltages $V$, we find that $m$ = 1 which can be attributed to direct tunneling (Fig. \ref{fig:IV}a, red line). We note that this regime is used to determine the junction resistances $R_{\text{JJ}}$ analyzed above. As the voltage is increasing, a transition occurs to a superlinear regime, which can be fitted by eq. 4 with $m$ = 2.75. As $m > $ 2, the dominant tunneling process is trap assisted (Fig. \ref{fig:IV}a, blue line), which indicates the presence of traps in the junction area \cite{Perkins_demonstration_2018}. The junction remains in this regime until breakthrough takes place at $V_{\text{BT}}$ = 1.2\,V, indicating a trap-related cause for the breakthrough. These traps are either microscopic defects located in the junction oxide, or macroscopic impurities on the junction interface. For this reason, we conducted a study with a reference wafer having similar junctions and analyzed their breakthrough distribution. The RT junction resistance spreads obtained for the reference and the qubit wafer are identical, which suggests that the junction of the two wafers are comparable. As the distributions of $V_{\text{BT}}$ for different junction sizes strongly overlap (Fig. \ref{fig:IV}b), with mean values for each size between 1.048\,V – 1.076\,V, we conclude that there is no significant dependence of $V_{\text{BT}}$ on junction area. Therefore the assumption of junction breakthrough due to defects is not verified, since we cannot observe a correlation to junction area. This observation is in accordance with reported critical defect densities causing a defect related breakthrough of $\sim 10^{-4}/$µm$^2$ for junctions fabricated with a similar process \cite{Lang_room_2025}. For junctions of our size ( $\sim$ 0.175\,µm$^2$), the amount of defects per JJ are negligible. We therefore assume a different, intrinsic breakthrough mechanism due to electronic traps in the junction oxide.
   
Interestingly, the RSD for $V_{\text{BT}}$ was measured to be $RSD_{V_{BT}}$ = 7.7\,\%, which is close the value of $RSD_{\text{RA}}$, discussed above. This indicates that the variability in oxide thickness and barrier height can be independently observed in the spread of the breakthrough voltage as well. It can be determined that the analysis of the breakthrough voltage and IV curve shape provides a significant advantage for studying JJs at RT.

\section{Cryogenic qubit characterization}
The following sections present the cryogenic characterization of superconducting qubits fabricated by the presented industry-grade fabrication approach and investigate correlations between the extensive RT, wafer-level characterization and low temperature properties of the devices. For this, individual chips of the design shown in Fig. \ref{fig:SC}b) are picked and wire bonded to cryogenic-compatible circuit boards (cryoboards) by aluminum wires before being mounted to a Bluefors LD400 dilution refrigerator. The LD400 cryostat is equipped with over 60 RF and 40 DC lines, with additional microwave switches implemented at the mixing chamber (MXC) stage to enable higher experimental throughput by the dynamical routing of signals. To determine the reliability and quality of the fabricated chips in terms of yield and performance indicators such as coherence time and frequency targeting, we picked devices from the full 200mm diameter of the wafer across a wide interval of RT resistance of $\sim$ 5.5 to 9.5\,k$\Omega$. 10 qubit chips with 4 qubits each, i.e. 40 qubits in total, were characterized.

\subsection{Frequency spectroscopy}

\begin{figure}[h]
    \centering
    \includegraphics[height=6cm]{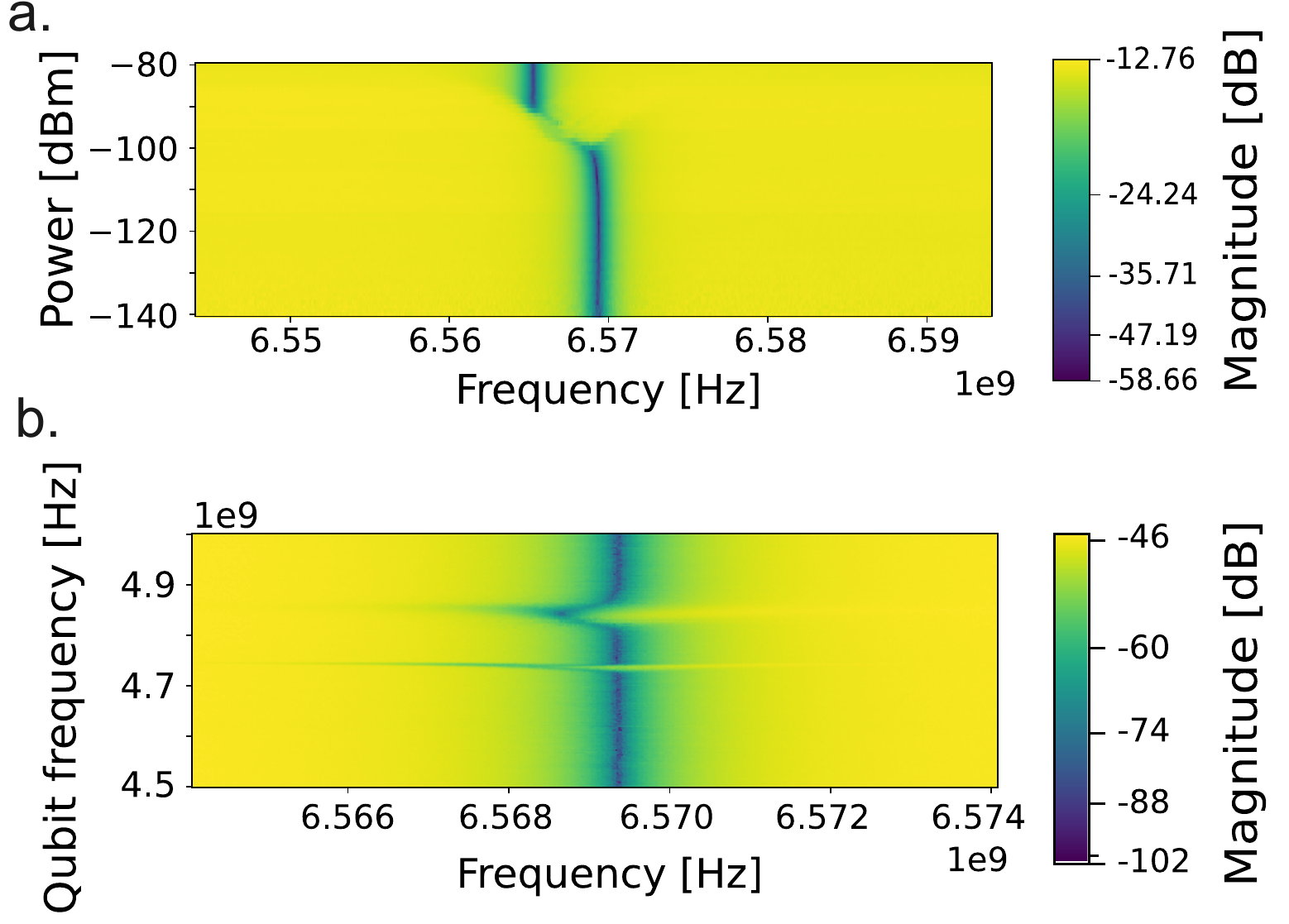}
    \caption{Example of a. one-tone and b. two-tone spectroscopy used to identify the resonance and qubit frequency and to detect qubit-resonator coupling.}
    \label{fig:12tone}
\end{figure} 

The qubit characterization was conducted in accordance with well-established methodologies \cite{naghiloo}, \cite{Krantz}. The qubits under test were addressed using read-out resonators. By determining the “punch-out” power, i.e. the readout power at which the induced current exceeds the critical current of the junction, one-tone spectroscopy (Fig. \ref{fig:12tone}a) serves as a first check whether an active qubit is coupled to the resonator. The strength of this coupling is indicated by magnitude of the shift. Two-tone spectroscopy (Fig. \ref{fig:12tone}b) was used to obtain values for the qubit frequencies and anharmonicities. We defined a qubit as functional, when clear signatures are observed in one-tone and two-tone spectroscopy. With this definition, we determine the yield of our functional qubits to be $>$ 92\,\%.

\subsection{Time domain characterization}

\begin{figure}[htbp]
    \centering
    \includegraphics[width=0.95\linewidth]{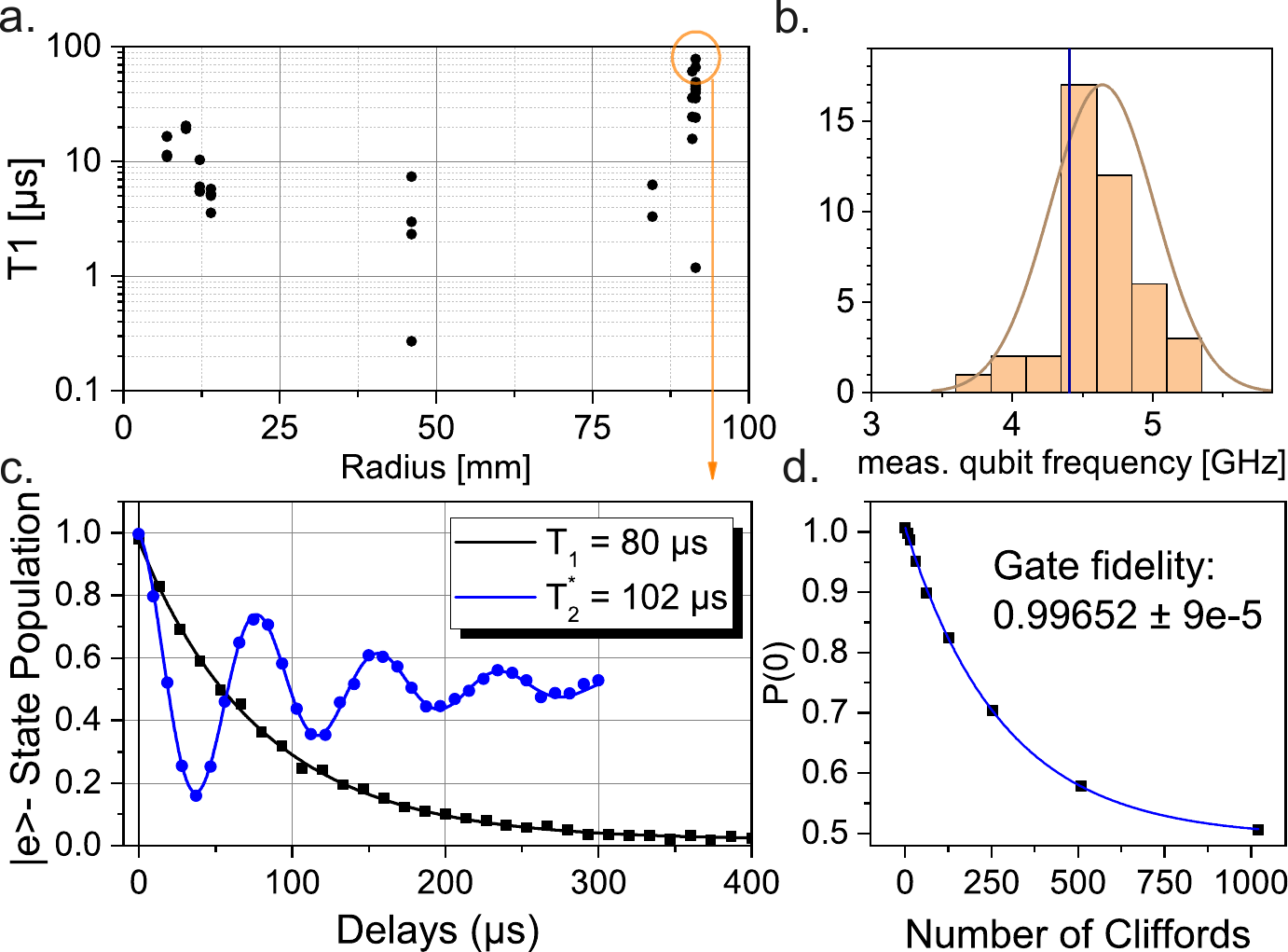}
    \caption{a. Measured $T_1$ vs. wafer radius, b. Qubit frequency distribution, c. $T_1$ and $T_2^*$ of best performing qubit and d. gate fidelity determined from randomized benchmarking.}
    \label{fig:qubit_char}
\end{figure}

Time-domain characterization of the superconducting qubits was performed using dedicated instrumentation from Zurich Instruments. A high-precision signal generator was employed to generate control pulses, while a Quantum Analyzer enabled accurate readout of the qubit states. Subsequent qubit calibration was performed using Rabi amplitude and Ramsey measurements. Following the calibration, the $T_1$ and $T_2^*$ times were determined, along with qubit state discrimination and single-qubit randomized benchmarking. Figure \ref{fig:qubit_char}a) shows the T1 times of the characterized qubits versus the on-wafer radius at which the respective chip was located. Across the full 200 mm wafer, we observe energy relaxation times in the microsecond regime. Together with the precise targeting of the design frequency (Fig. \ref{fig:qubit_char}b) ) these results underline the potential of our industry-grade, wafer-level fabrication approach to yield high quality superconductive qubits while maintaining large area uniformity. While picking across the full wafer radius, the total frequency spread of the characterized qubits lies at $\approx$ 8.4\,\%, providing a promising outlook for reaching the necessary frequency targeting in large QPU devices. By precise room-temperature characterization and chip pre-selection, the deviation from the target and the overall spread of the frequency of a set of characterized qubits can be further optimized.
For the best qubits, the $T_1$ reaches 80\,µs and the $T_2^*$ $>$ 100\,µs as shown in Fig. \ref{fig:qubit_char}c) and hence comparable values to previously reported results for chips fabricated with a similar processing approach \cite{VanDamme_Advanced_2024} are observed. The quality of our qubits is also reflected by reaching single-qubit gate fidelities above 99.6\% (Fig. \ref{fig:qubit_char}d), which were obtained by randomized benchmarking experiments.

\begin{figure}[h!]
    \centering
    \includegraphics[width=0.95\linewidth, clip]{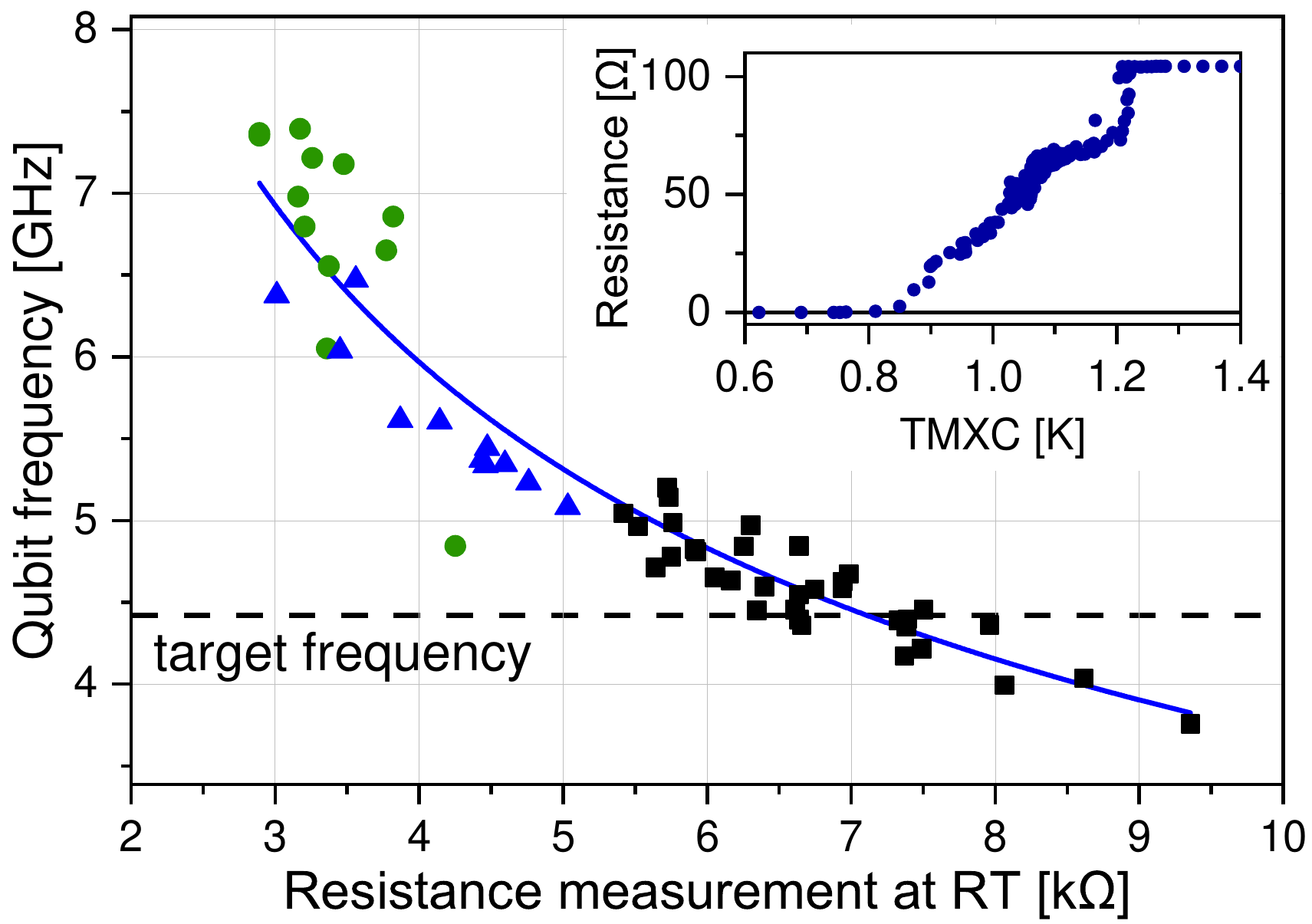}    
    \caption{RT resistivity measurements vs. qubit frequencies measured at 10\,mK from different qubit runs (green dots, blue triangles and black squares), fitted with an Ambegaokar-Baratoff model using a fitting parameter of $T_c$ $\sim$ 0.71\,K. $T_c$ was verified by measuring the resistance of a short-circuited 0.35\,µm junction top electrode during cooldown (nested figure).}
    \label{fig:BA}
\end{figure} 

To connect the wafer-level measurements at RT with the cryostatic qubit characterization, Fig. \ref{fig:BA} plots the normal state resistance $R_n$ of a qubit’s Josephson Junction versus its frequency $f_{\text{01}}$ obtained by two-tone measurements at 10\,mK. To confirm the reproducibility of the results, the plot includes the qubits presented in Fig. \ref{fig:qubit_char} (black squares) and additional data of two test wafers from independent fabrication runs (green dots and blue triangles), where changes to the Al etch chemistry lead to a shift in normal state resistance. This also spans the data set to a broader range for more accurate fitting applying the Ambegaokar-Baratoff model \cite{ambegaokar_tunneling_1963}
\begin{equation}
    f_{\text{01}} = \sqrt{\frac{1}{R_n}\frac{0.882k_BT_c}{hC_q}}-\frac{e^2}{2hC_q}
\end{equation}
with the Boltzmann constant $k_B$, Planck constant $h$, the electron charge $e$, and the qubit capacitance $C_q$ being set constant to the design value of 86\,fF, which was verified by evaluating the anharmonicity of the qubits. Leaving the critical temperature as a free fit parameter, a value of T$_c\approx0.71$\,K is obtained, posing an offset of the curve from the expected literature value of 1.2\,K for aluminum. The inset of Fig. \ref{fig:BA} shows a four-point DC measurement of an aluminum short on a test device with a channel width of 0.35\,µm corresponding to the minimal dimension in our qubit device (width of bottom electrode). The measurement shows the onset of superconductivity at the expected 1.2\,K, but full superconductivity is only reached at a lower temperature of 0.82\,K, which could explain the observed offset in the Ambegaokar-Baratoff relation shown in the main figure (Fig. \ref{fig:BA}). Furthermore, the shift could also origin from an effect offsetting the normal state resistance of the JJ like e.g. leakage currents through the substrate or other additional parasitic resistances within the current path. However, since this offset is stable between processing runs, the model can be used to predict the qubit frequency from the RT resistance measurement and hence poses a powerful tool towards precise frequency targeting by applying proper pre-selection of qubit chips.

\section{Conclusion}
In conclusion, this study presents a comprehensive methodology for the fabrication and characterization of superconducting qubits using a CMOS-compatible process on a 200 mm wafer. By integrating process control monitoring through the use of test structures, we successfully assessed the electrical characteristics and variability of qubit Josephson junctions at RT, providing critical insights before cryogenic measurements. The analysis of shorts enabled an efficient evaluation of the metal structuring variability, leading to area variation of $RSD_\text{A}$ = 10.1\,\% and the calculated offset between design and fabrication width up to 99.3 nm. In combination with the junctions analysis of the resistance variation of $RSD_{\text{JJ}}$ = 12.4\,\%, we could calculate the oxide barrier variation to be $RSD_{\text{RA}}$ = 7.2\,\%. Interestingly, we found that the breakthrough voltage has a similar across-wafer variation as the oxide barrier, making it highly important for detailed qubit RT characterization. In total, we achieved a yield of 92.8\% for functional junctions across the 200 mm wafer. 
The cryogenic characterization demonstrated the effective targeting of qubit frequencies, with $T_1$ times reaching up to 80\,µs and $T_2^*$ values exceeding 100\,µs, indicative of high-quality devices. With the Ambegaokar-Baratoff formular applying to all our samples of several fabrication cycles with a $T_c$ $\sim$ 0.71\,K, optimized frequency targeting was enabled. Our results highlight the robustness and scalability of the fabrication process, positioning it as a viable pathway for future advancements in quantum processing units, particularly in the context of 3D integrated multi-chip systems \cite{Niu_low_2023, smith_scaling_2022, Field_Modular_2024}. Overall, this work lays the groundwork for the continued development and implementation of superconducting qubits in quantum computing applications.

\section*{Acknowledgements}
The authors acknowledge helpful discussions with G. Huber, I. Tsitsilin, F. Haslbeck, C. Schneider, L. Koch and N. Bruckmoser from the Quantum Computing group at the Walther Meissner Institute. We would also like to thank Tianmu Zhang and Lukas Sigl form Zurich Instruments for their support in setting up the qubit measurement system. 
Moreover, we would like to thank Dr. Lars Nebrich for his support in layouting. We also appreciate Martin Heigl and Martin König for their help and valuable discussions in process development and the whole Fraunhofer EMFT clean room staff for the professional fabrication. We would also like to thank Samuel Taubenberger for supporting us in setting up the measurement software.
\\ \\
This work was funded by the Munich Quantum Valley (MQV) – Consortium Scalable Hardware and Systems Engineering (SHARE), funded by the Bavarian State Government with funds from the Hightech Agenda Bavaria, the Munich Quantum Valley Quantum Computer Demonstrator - Superconducting Qubits (MUNIQC-SC) 13N16188, funded by the Federal Ministry of Education and Research, Germany, and the Open Superconducting Quantum Computers (OpenSuperQPlus) Project - European Quantum Technology Flagship.

\bibliography{paper.bib}

\begin{thebibliography}{21}%
\makeatletter
\providecommand \@ifxundefined [1]{%
 \@ifx{#1\undefined}
}%
\providecommand \@ifnum [1]{%
 \ifnum #1\expandafter \@firstoftwo
 \else \expandafter \@secondoftwo
 \fi
}%
\providecommand \@ifx [1]{%
 \ifx #1\expandafter \@firstoftwo
 \else \expandafter \@secondoftwo
 \fi
}%
\providecommand \natexlab [1]{#1}%
\providecommand \enquote  [1]{``#1''}%
\providecommand \bibnamefont  [1]{#1}%
\providecommand \bibfnamefont [1]{#1}%
\providecommand \citenamefont [1]{#1}%
\providecommand \href@noop [0]{\@secondoftwo}%
\providecommand \href [0]{\begingroup \@sanitize@url \@href}%
\providecommand \@href[1]{\@@startlink{#1}\@@href}%
\providecommand \@@href[1]{\endgroup#1\@@endlink}%
\providecommand \@sanitize@url [0]{\catcode `\\12\catcode `\$12\catcode
  `\&12\catcode `\#12\catcode `\^12\catcode `\_12\catcode `\%12\relax}%
\providecommand \@@startlink[1]{}%
\providecommand \@@endlink[0]{}%
\providecommand \url  [0]{\begingroup\@sanitize@url \@url }%
\providecommand \@url [1]{\endgroup\@href {#1}{\urlprefix }}%
\providecommand \urlprefix  [0]{URL }%
\providecommand \Eprint [0]{\href }%
\providecommand \doibase [0]{https://doi.org/}%
\providecommand \selectlanguage [0]{\@gobble}%
\providecommand \bibinfo  [0]{\@secondoftwo}%
\providecommand \bibfield  [0]{\@secondoftwo}%
\providecommand \translation [1]{[#1]}%
\providecommand \BibitemOpen [0]{}%
\providecommand \bibitemStop [0]{}%
\providecommand \bibitemNoStop [0]{.\EOS\space}%
\providecommand \EOS [0]{\spacefactor3000\relax}%
\providecommand \BibitemShut  [1]{\csname bibitem#1\endcsname}%
\let\auto@bib@innerbib\@empty
\bibitem [{\citenamefont {Mohseni}\ \emph {et~al.}(2025)\citenamefont
  {Mohseni}, \citenamefont {Scherer}, \citenamefont {Johnson}, \citenamefont
  {Wertheim}, \citenamefont {Otten}, \citenamefont {Aadit}, \citenamefont
  {Alexeev}, \citenamefont {Bresniker}, \citenamefont {Camsari}, \citenamefont
  {Chapman}, \citenamefont {Chatterjee}, \citenamefont {Dagnew}, \citenamefont
  {Esposito}, \citenamefont {Fahim}, \citenamefont {Fiorentino}, \citenamefont
  {Gajjar}, \citenamefont {Khalid}, \citenamefont {Kong}, \citenamefont
  {Kulchytskyy}, \citenamefont {Kyoseva}, \citenamefont {Li}, \citenamefont
  {Lott}, \citenamefont {Markov}, \citenamefont {McDermott}, \citenamefont
  {Pedretti}, \citenamefont {Rao}, \citenamefont {Rieffel}, \citenamefont
  {Silva}, \citenamefont {Sorebo}, \citenamefont {Spentzouris}, \citenamefont
  {Steiner}, \citenamefont {Torosov}, \citenamefont {Venturelli}, \citenamefont
  {Visser}, \citenamefont {Webb}, \citenamefont {Zhan}, \citenamefont {Cohen},
  \citenamefont {Ronagh}, \citenamefont {Ho}, \citenamefont {Beausoleil},\ and\
  \citenamefont {Martinis}}]{mohseni_How_2025}%
  \BibitemOpen
  \bibfield  {author} {\bibinfo {author} {\bibfnamefont {M.}~\bibnamefont
  {Mohseni}}, \bibinfo {author} {\bibfnamefont {A.}~\bibnamefont {Scherer}},
  \bibinfo {author} {\bibfnamefont {K.~G.}\ \bibnamefont {Johnson}}, \bibinfo
  {author} {\bibfnamefont {O.}~\bibnamefont {Wertheim}}, \bibinfo {author}
  {\bibfnamefont {M.}~\bibnamefont {Otten}}, \bibinfo {author} {\bibfnamefont
  {N.~A.}\ \bibnamefont {Aadit}}, \bibinfo {author} {\bibfnamefont
  {Y.}~\bibnamefont {Alexeev}}, \bibinfo {author} {\bibfnamefont {K.~M.}\
  \bibnamefont {Bresniker}}, \bibinfo {author} {\bibfnamefont {K.~Y.}\
  \bibnamefont {Camsari}}, \bibinfo {author} {\bibfnamefont {B.}~\bibnamefont
  {Chapman}}, \bibinfo {author} {\bibfnamefont {S.}~\bibnamefont {Chatterjee}},
  \bibinfo {author} {\bibfnamefont {G.~A.}\ \bibnamefont {Dagnew}}, \bibinfo
  {author} {\bibfnamefont {A.}~\bibnamefont {Esposito}}, \bibinfo {author}
  {\bibfnamefont {F.}~\bibnamefont {Fahim}}, \bibinfo {author} {\bibfnamefont
  {M.}~\bibnamefont {Fiorentino}}, \bibinfo {author} {\bibfnamefont
  {A.}~\bibnamefont {Gajjar}}, \bibinfo {author} {\bibfnamefont
  {A.}~\bibnamefont {Khalid}}, \bibinfo {author} {\bibfnamefont
  {X.}~\bibnamefont {Kong}}, \bibinfo {author} {\bibfnamefont {B.}~\bibnamefont
  {Kulchytskyy}}, \bibinfo {author} {\bibfnamefont {E.}~\bibnamefont
  {Kyoseva}}, \bibinfo {author} {\bibfnamefont {R.}~\bibnamefont {Li}},
  \bibinfo {author} {\bibfnamefont {P.~A.}\ \bibnamefont {Lott}}, \bibinfo
  {author} {\bibfnamefont {I.~L.}\ \bibnamefont {Markov}}, \bibinfo {author}
  {\bibfnamefont {R.~F.}\ \bibnamefont {McDermott}}, \bibinfo {author}
  {\bibfnamefont {G.}~\bibnamefont {Pedretti}}, \bibinfo {author}
  {\bibfnamefont {P.}~\bibnamefont {Rao}}, \bibinfo {author} {\bibfnamefont
  {E.}~\bibnamefont {Rieffel}}, \bibinfo {author} {\bibfnamefont
  {A.}~\bibnamefont {Silva}}, \bibinfo {author} {\bibfnamefont
  {J.}~\bibnamefont {Sorebo}}, \bibinfo {author} {\bibfnamefont
  {P.}~\bibnamefont {Spentzouris}}, \bibinfo {author} {\bibfnamefont
  {Z.}~\bibnamefont {Steiner}}, \bibinfo {author} {\bibfnamefont
  {B.}~\bibnamefont {Torosov}}, \bibinfo {author} {\bibfnamefont
  {D.}~\bibnamefont {Venturelli}}, \bibinfo {author} {\bibfnamefont {R.~J.}\
  \bibnamefont {Visser}}, \bibinfo {author} {\bibfnamefont {Z.}~\bibnamefont
  {Webb}}, \bibinfo {author} {\bibfnamefont {X.}~\bibnamefont {Zhan}}, \bibinfo
  {author} {\bibfnamefont {Y.}~\bibnamefont {Cohen}}, \bibinfo {author}
  {\bibfnamefont {P.}~\bibnamefont {Ronagh}}, \bibinfo {author} {\bibfnamefont
  {A.}~\bibnamefont {Ho}}, \bibinfo {author} {\bibfnamefont {R.~G.}\
  \bibnamefont {Beausoleil}},\ and\ \bibinfo {author} {\bibfnamefont {J.~M.}\
  \bibnamefont {Martinis}},\ }\href {https://arxiv.org/abs/2411.10406}
  {\bibinfo {title} {How to build a quantum supercomputer: Scaling from
  hundreds to millions of qubits}} (\bibinfo {year} {2025}),\ \Eprint
  {https://arxiv.org/abs/2411.10406} {arXiv:2411.10406 [quant-ph]} \BibitemShut
  {NoStop}%
\bibitem [{\citenamefont {Mayer}\ \emph {et~al.}(2025)\citenamefont {Mayer},
  \citenamefont {Bender}, \citenamefont {Lang}, \citenamefont {Luo},
  \citenamefont {Weber}, \citenamefont {Guizan}, \citenamefont {Dhieb},
  \citenamefont {Zahn}, \citenamefont {Schwarzenbach}, \citenamefont {Hell},
  \citenamefont {Andronic}, \citenamefont {Drost}, \citenamefont {Neumeier},
  \citenamefont {Lerch}, \citenamefont {Nebrich}, \citenamefont {Hagelauer},
  \citenamefont {Eisele}, \citenamefont {Pereira},\ and\ \citenamefont
  {Kutter}}]{Mayer_3D_2025}%
  \BibitemOpen
  \bibfield  {author} {\bibinfo {author} {\bibfnamefont {T.}~\bibnamefont
  {Mayer}}, \bibinfo {author} {\bibfnamefont {H.}~\bibnamefont {Bender}},
  \bibinfo {author} {\bibfnamefont {S.~J.~K.}\ \bibnamefont {Lang}}, \bibinfo
  {author} {\bibfnamefont {Z.}~\bibnamefont {Luo}}, \bibinfo {author}
  {\bibfnamefont {J.}~\bibnamefont {Weber}}, \bibinfo {author} {\bibfnamefont
  {C.~M.}\ \bibnamefont {Guizan}}, \bibinfo {author} {\bibfnamefont
  {C.}~\bibnamefont {Dhieb}}, \bibinfo {author} {\bibfnamefont
  {D.}~\bibnamefont {Zahn}}, \bibinfo {author} {\bibfnamefont {L.}~\bibnamefont
  {Schwarzenbach}}, \bibinfo {author} {\bibfnamefont {W.}~\bibnamefont {Hell}},
  \bibinfo {author} {\bibfnamefont {M.}~\bibnamefont {Andronic}}, \bibinfo
  {author} {\bibfnamefont {A.}~\bibnamefont {Drost}}, \bibinfo {author}
  {\bibfnamefont {K.}~\bibnamefont {Neumeier}}, \bibinfo {author}
  {\bibfnamefont {W.}~\bibnamefont {Lerch}}, \bibinfo {author} {\bibfnamefont
  {L.}~\bibnamefont {Nebrich}}, \bibinfo {author} {\bibfnamefont
  {A.}~\bibnamefont {Hagelauer}}, \bibinfo {author} {\bibfnamefont
  {I.}~\bibnamefont {Eisele}}, \bibinfo {author} {\bibfnamefont {R.~N.}\
  \bibnamefont {Pereira}},\ and\ \bibinfo {author} {\bibfnamefont
  {C.}~\bibnamefont {Kutter}},\ }\href {https://arxiv.org/abs/2505.04337}
  {\bibinfo {title} {3d-integrated superconducting qubits: Cmos-compatible,
  wafer-scale processing for flip-chip architectures}} (\bibinfo {year}
  {2025}),\ \Eprint {https://arxiv.org/abs/2505.04337} {arXiv:2505.04337
  [quant-ph]} \BibitemShut {NoStop}%
\bibitem [{\citenamefont {Rosenberg}\ \emph {et~al.}(2017)\citenamefont
  {Rosenberg}, \citenamefont {Kim}, \citenamefont {Das}, \citenamefont {Yost},
  \citenamefont {Gustavsson}, \citenamefont {Hover}, \citenamefont {Krantz},
  \citenamefont {Melville}, \citenamefont {Racz}, \citenamefont {Samach},
  \citenamefont {Weber}, \citenamefont {Yan}, \citenamefont {Yoder},
  \citenamefont {Kerman},\ and\ \citenamefont {Oliver}}]{Rosenberg_3D_2017}%
  \BibitemOpen
  \bibfield  {author} {\bibinfo {author} {\bibfnamefont {D.}~\bibnamefont
  {Rosenberg}}, \bibinfo {author} {\bibfnamefont {D.}~\bibnamefont {Kim}},
  \bibinfo {author} {\bibfnamefont {R.}~\bibnamefont {Das}}, \bibinfo {author}
  {\bibfnamefont {D.}~\bibnamefont {Yost}}, \bibinfo {author} {\bibfnamefont
  {S.}~\bibnamefont {Gustavsson}}, \bibinfo {author} {\bibfnamefont
  {D.}~\bibnamefont {Hover}}, \bibinfo {author} {\bibfnamefont
  {P.}~\bibnamefont {Krantz}}, \bibinfo {author} {\bibfnamefont
  {A.}~\bibnamefont {Melville}}, \bibinfo {author} {\bibfnamefont
  {L.}~\bibnamefont {Racz}}, \bibinfo {author} {\bibfnamefont {G.~O.}\
  \bibnamefont {Samach}}, \bibinfo {author} {\bibfnamefont {S.~J.}\
  \bibnamefont {Weber}}, \bibinfo {author} {\bibfnamefont {F.}~\bibnamefont
  {Yan}}, \bibinfo {author} {\bibfnamefont {J.~L.}\ \bibnamefont {Yoder}},
  \bibinfo {author} {\bibfnamefont {A.~J.}\ \bibnamefont {Kerman}},\ and\
  \bibinfo {author} {\bibfnamefont {W.~D.}\ \bibnamefont {Oliver}},\ }\bibfield
   {title} {\bibinfo {title} {3d integrated superconducting qubits},\
  }\bibfield  {journal} {\bibinfo  {journal} {npj Quantum Information}\
  }\textbf {\bibinfo {volume} {3}},\ \href
  {https://doi.org/10.1038/s41534-017-0044-0} {10.1038/s41534-017-0044-0}
  (\bibinfo {year} {2017})\BibitemShut {NoStop}%
\bibitem [{\citenamefont {Van~Damme}\ \emph {et~al.}(2024)\citenamefont
  {Van~Damme}, \citenamefont {Massar}, \citenamefont {Acharya}, \citenamefont
  {Ivanov}, \citenamefont {Perez~Lozano}, \citenamefont {Canvel}, \citenamefont
  {Demarets}, \citenamefont {Vangoidsenhoven}, \citenamefont {Hermans},
  \citenamefont {Lai}, \citenamefont {Vadiraj}, \citenamefont {Mongillo},
  \citenamefont {Wan}, \citenamefont {De~Boeck}, \citenamefont {Potočnik},\
  and\ \citenamefont {De~Greve}}]{VanDamme_Advanced_2024}%
  \BibitemOpen
  \bibfield  {author} {\bibinfo {author} {\bibfnamefont {J.}~\bibnamefont
  {Van~Damme}}, \bibinfo {author} {\bibfnamefont {S.}~\bibnamefont {Massar}},
  \bibinfo {author} {\bibfnamefont {R.}~\bibnamefont {Acharya}}, \bibinfo
  {author} {\bibfnamefont {T.}~\bibnamefont {Ivanov}}, \bibinfo {author}
  {\bibfnamefont {D.}~\bibnamefont {Perez~Lozano}}, \bibinfo {author}
  {\bibfnamefont {Y.}~\bibnamefont {Canvel}}, \bibinfo {author} {\bibfnamefont
  {M.}~\bibnamefont {Demarets}}, \bibinfo {author} {\bibfnamefont
  {D.}~\bibnamefont {Vangoidsenhoven}}, \bibinfo {author} {\bibfnamefont
  {Y.}~\bibnamefont {Hermans}}, \bibinfo {author} {\bibfnamefont {J.~G.}\
  \bibnamefont {Lai}}, \bibinfo {author} {\bibfnamefont {A.~M.}\ \bibnamefont
  {Vadiraj}}, \bibinfo {author} {\bibfnamefont {M.}~\bibnamefont {Mongillo}},
  \bibinfo {author} {\bibfnamefont {D.}~\bibnamefont {Wan}}, \bibinfo {author}
  {\bibfnamefont {J.}~\bibnamefont {De~Boeck}}, \bibinfo {author}
  {\bibfnamefont {A.}~\bibnamefont {Potočnik}},\ and\ \bibinfo {author}
  {\bibfnamefont {K.}~\bibnamefont {De~Greve}},\ }\bibfield  {title} {\bibinfo
  {title} {Advanced {CMOS} manufacturing of superconducting qubits on 300 mm
  wafers},\ }\href {https://doi.org/10.1038/s41586-024-07941-9} {\bibfield
  {journal} {\bibinfo  {journal} {Nature}\ }\textbf {\bibinfo {volume} {634}},\
  \bibinfo {pages} {74} (\bibinfo {year} {2024})}\BibitemShut {NoStop}%
\bibitem [{\citenamefont {Stehli}\ \emph {et~al.}(2020)\citenamefont {Stehli},
  \citenamefont {Brehm}, \citenamefont {Wolz}, \citenamefont {Baity},
  \citenamefont {Danilin}, \citenamefont {Seferai}, \citenamefont {Rotzinger},
  \citenamefont {Ustinov},\ and\ \citenamefont
  {Weides}}]{Stehil_Coherent_2020}%
  \BibitemOpen
  \bibfield  {author} {\bibinfo {author} {\bibfnamefont {A.}~\bibnamefont
  {Stehli}}, \bibinfo {author} {\bibfnamefont {J.~D.}\ \bibnamefont {Brehm}},
  \bibinfo {author} {\bibfnamefont {T.}~\bibnamefont {Wolz}}, \bibinfo {author}
  {\bibfnamefont {P.}~\bibnamefont {Baity}}, \bibinfo {author} {\bibfnamefont
  {S.}~\bibnamefont {Danilin}}, \bibinfo {author} {\bibfnamefont
  {V.}~\bibnamefont {Seferai}}, \bibinfo {author} {\bibfnamefont
  {H.}~\bibnamefont {Rotzinger}}, \bibinfo {author} {\bibfnamefont {A.~V.}\
  \bibnamefont {Ustinov}},\ and\ \bibinfo {author} {\bibfnamefont
  {M.}~\bibnamefont {Weides}},\ }\bibfield  {title} {\bibinfo {title} {Coherent
  superconducting qubits from a subtractive junction fabrication process},\
  }\href {https://doi.org/10.1063/5.0023533} {\bibfield  {journal} {\bibinfo
  {journal} {Applied Physics Letters}\ }\textbf {\bibinfo {volume} {117}},\
  \bibinfo {pages} {124005} (\bibinfo {year} {2020})},\ \Eprint
  {https://arxiv.org/abs/https://pubs.aip.org/aip/apl/article-pdf/doi/10.1063/5.0023533/14539624/124005\_1\_online.pdf}
  {https://pubs.aip.org/aip/apl/article-pdf/doi/10.1063/5.0023533/14539624/124005\_1\_online.pdf}
  \BibitemShut {NoStop}%
\bibitem [{\citenamefont {Lang}\ \emph {et~al.}(2023)\citenamefont {Lang},
  \citenamefont {Schewski}, \citenamefont {Eisele}, \citenamefont {Kutter},\
  and\ \citenamefont {Lerch}}]{lang_aluminum_2023}%
  \BibitemOpen
  \bibfield  {author} {\bibinfo {author} {\bibfnamefont {S.}~\bibnamefont
  {Lang}}, \bibinfo {author} {\bibfnamefont {A.}~\bibnamefont {Schewski}},
  \bibinfo {author} {\bibfnamefont {I.}~\bibnamefont {Eisele}}, \bibinfo
  {author} {\bibfnamefont {C.}~\bibnamefont {Kutter}},\ and\ \bibinfo {author}
  {\bibfnamefont {W.}~\bibnamefont {Lerch}},\ }\bibfield  {title} {\bibinfo
  {title} {Aluminum {Josephson} {Junction} {Formation} on 200mm {Wafers}
  {Using} {Different} {Oxidation} {Techniques}},\ }\href
  {https://doi.org/10.1149/11101.0041ecst} {\bibfield  {journal} {\bibinfo
  {journal} {ECS Transactions}\ }\textbf {\bibinfo {volume} {111}},\ \bibinfo
  {pages} {41} (\bibinfo {year} {2023})}\BibitemShut {NoStop}%
\bibitem [{\citenamefont {Wan}\ \emph {et~al.}(2021)\citenamefont {Wan},
  \citenamefont {Couet}, \citenamefont {Piao}, \citenamefont {Souriau},
  \citenamefont {Canvel}, \citenamefont {Tsvetanova}, \citenamefont
  {Vangoidsenhoven}, \citenamefont {Thiam}, \citenamefont {Pacco},
  \citenamefont {Potočnik}, \citenamefont {Mongillo}, \citenamefont {Ivanov},
  \citenamefont {Jussot}, \citenamefont {Verjauw}, \citenamefont {Acharya},
  \citenamefont {Lazzarino}, \citenamefont {Govoreanu},\ and\ \citenamefont
  {Radu}}]{Wan_Fabrication_2021}%
  \BibitemOpen
  \bibfield  {author} {\bibinfo {author} {\bibfnamefont {D.}~\bibnamefont
  {Wan}}, \bibinfo {author} {\bibfnamefont {S.}~\bibnamefont {Couet}}, \bibinfo
  {author} {\bibfnamefont {X.}~\bibnamefont {Piao}}, \bibinfo {author}
  {\bibfnamefont {L.}~\bibnamefont {Souriau}}, \bibinfo {author} {\bibfnamefont
  {Y.}~\bibnamefont {Canvel}}, \bibinfo {author} {\bibfnamefont
  {D.}~\bibnamefont {Tsvetanova}}, \bibinfo {author} {\bibfnamefont
  {D.}~\bibnamefont {Vangoidsenhoven}}, \bibinfo {author} {\bibfnamefont
  {A.}~\bibnamefont {Thiam}}, \bibinfo {author} {\bibfnamefont
  {A.}~\bibnamefont {Pacco}}, \bibinfo {author} {\bibfnamefont
  {A.}~\bibnamefont {Potočnik}}, \bibinfo {author} {\bibfnamefont
  {M.}~\bibnamefont {Mongillo}}, \bibinfo {author} {\bibfnamefont
  {T.}~\bibnamefont {Ivanov}}, \bibinfo {author} {\bibfnamefont
  {J.}~\bibnamefont {Jussot}}, \bibinfo {author} {\bibfnamefont
  {J.}~\bibnamefont {Verjauw}}, \bibinfo {author} {\bibfnamefont
  {R.}~\bibnamefont {Acharya}}, \bibinfo {author} {\bibfnamefont
  {F.}~\bibnamefont {Lazzarino}}, \bibinfo {author} {\bibfnamefont
  {B.}~\bibnamefont {Govoreanu}},\ and\ \bibinfo {author} {\bibfnamefont
  {I.~P.}\ \bibnamefont {Radu}},\ }\bibfield  {title} {\bibinfo {title}
  {Fabrication and room temperature characterization of trilayer junctions for
  the development of superconducting qubits on 300 mm wafers},\ }\href
  {https://doi.org/10.35848/1347-4065/abe5bb} {\bibfield  {journal} {\bibinfo
  {journal} {Japanese Journal of Applied Physics}\ }\textbf {\bibinfo {volume}
  {60}},\ \bibinfo {pages} {SBBI04} (\bibinfo {year} {2021})}\BibitemShut
  {NoStop}%
\bibitem [{\citenamefont {Lang}\ \emph {et~al.}(2025)\citenamefont {Lang},
  \citenamefont {Eisele}, \citenamefont {Weber}, \citenamefont {Schewski},
  \citenamefont {Lerch}, \citenamefont {Pereira},\ and\ \citenamefont
  {Kutter}}]{Lang_room_2025}%
  \BibitemOpen
  \bibfield  {author} {\bibinfo {author} {\bibfnamefont {S.~J.~K.}\
  \bibnamefont {Lang}}, \bibinfo {author} {\bibfnamefont {I.}~\bibnamefont
  {Eisele}}, \bibinfo {author} {\bibfnamefont {J.}~\bibnamefont {Weber}},
  \bibinfo {author} {\bibfnamefont {A.}~\bibnamefont {Schewski}}, \bibinfo
  {author} {\bibfnamefont {W.}~\bibnamefont {Lerch}}, \bibinfo {author}
  {\bibfnamefont {R.~N.}\ \bibnamefont {Pereira}},\ and\ \bibinfo {author}
  {\bibfnamefont {C.}~\bibnamefont {Kutter}},\ }\href
  {https://arxiv.org/abs/2504.16686} {\bibinfo {title} {Wafer-scale
  characterization of al/alxoy/al josephson junctions at room temperature}}
  (\bibinfo {year} {2025}),\ \Eprint {https://arxiv.org/abs/2504.16686}
  {arXiv:2504.16686 [quant-ph]} \BibitemShut {NoStop}%
\bibitem [{\citenamefont {Pishchimova}\ \emph {et~al.}(2022)\citenamefont
  {Pishchimova}, \citenamefont {Smirnov}, \citenamefont {Ezenkova},
  \citenamefont {Krivko}, \citenamefont {Zikiy}, \citenamefont {Moskalev},
  \citenamefont {Ivanov}, \citenamefont {Korshakov},\ and\ \citenamefont
  {Rodionov}}]{pishchimova_improving_2022}%
  \BibitemOpen
  \bibfield  {author} {\bibinfo {author} {\bibfnamefont {A.~A.}\ \bibnamefont
  {Pishchimova}}, \bibinfo {author} {\bibfnamefont {N.~S.}\ \bibnamefont
  {Smirnov}}, \bibinfo {author} {\bibfnamefont {D.~A.}\ \bibnamefont
  {Ezenkova}}, \bibinfo {author} {\bibfnamefont {E.~A.}\ \bibnamefont
  {Krivko}}, \bibinfo {author} {\bibfnamefont {E.~V.}\ \bibnamefont {Zikiy}},
  \bibinfo {author} {\bibfnamefont {D.~O.}\ \bibnamefont {Moskalev}}, \bibinfo
  {author} {\bibfnamefont {A.~I.}\ \bibnamefont {Ivanov}}, \bibinfo {author}
  {\bibfnamefont {N.~D.}\ \bibnamefont {Korshakov}},\ and\ \bibinfo {author}
  {\bibfnamefont {I.~A.}\ \bibnamefont {Rodionov}},\ }\href
  {https://arxiv.org/abs/2210.15293} {\bibinfo {title} {Improving josephson
  junction reproducibility for superconducting quantum circuits: junction area
  fluctuation}} (\bibinfo {year} {2022}),\ \Eprint
  {https://arxiv.org/abs/2210.15293} {arXiv:2210.15293 [quant-ph]} \BibitemShut
  {NoStop}%
\bibitem [{\citenamefont {Osman}\ \emph {et~al.}(2021)\citenamefont {Osman},
  \citenamefont {Simon}, \citenamefont {Bengtsson}, \citenamefont {Kosen},
  \citenamefont {Krantz}, \citenamefont {P.~Lozano}, \citenamefont
  {Scigliuzzo}, \citenamefont {Delsing}, \citenamefont {Bylander},\ and\
  \citenamefont {Fadavi~Roudsari}}]{Osman_Simplified_2021}%
  \BibitemOpen
  \bibfield  {author} {\bibinfo {author} {\bibfnamefont {A.}~\bibnamefont
  {Osman}}, \bibinfo {author} {\bibfnamefont {J.}~\bibnamefont {Simon}},
  \bibinfo {author} {\bibfnamefont {A.}~\bibnamefont {Bengtsson}}, \bibinfo
  {author} {\bibfnamefont {S.}~\bibnamefont {Kosen}}, \bibinfo {author}
  {\bibfnamefont {P.}~\bibnamefont {Krantz}}, \bibinfo {author} {\bibfnamefont
  {D.}~\bibnamefont {P.~Lozano}}, \bibinfo {author} {\bibfnamefont
  {M.}~\bibnamefont {Scigliuzzo}}, \bibinfo {author} {\bibfnamefont
  {P.}~\bibnamefont {Delsing}}, \bibinfo {author} {\bibfnamefont
  {J.}~\bibnamefont {Bylander}},\ and\ \bibinfo {author} {\bibfnamefont
  {A.}~\bibnamefont {Fadavi~Roudsari}},\ }\bibfield  {title} {\bibinfo {title}
  {Simplified josephson-junction fabrication process for reproducibly
  high-performance superconducting qubits},\ }\bibfield  {journal} {\bibinfo
  {journal} {Applied Physics Letters}\ }\textbf {\bibinfo {volume} {118}},\
  \href {https://doi.org/10.1063/5.0037093} {10.1063/5.0037093} (\bibinfo
  {year} {2021})\BibitemShut {NoStop}%
\bibitem [{\citenamefont {Siddiqi}(2021)}]{siddiqi_engineering_2021}%
  \BibitemOpen
  \bibfield  {author} {\bibinfo {author} {\bibfnamefont {I.}~\bibnamefont
  {Siddiqi}},\ }\bibfield  {title} {\bibinfo {title} {Engineering
  high-coherence superconducting qubits},\ }\href
  {https://doi.org/10.1038/s41578-021-00370-4} {\bibfield  {journal} {\bibinfo
  {journal} {Nature Reviews Materials}\ }\textbf {\bibinfo {volume} {6}},\
  \bibinfo {pages} {875} (\bibinfo {year} {2021})}\BibitemShut {NoStop}%
\bibitem [{\citenamefont {Wu}\ \emph {et~al.}(2017)\citenamefont {Wu},
  \citenamefont {Long}, \citenamefont {Ku}, \citenamefont {Lake}, \citenamefont
  {Bal},\ and\ \citenamefont {Pappas}}]{Wu_Overlap_2017}%
  \BibitemOpen
  \bibfield  {author} {\bibinfo {author} {\bibfnamefont {X.}~\bibnamefont
  {Wu}}, \bibinfo {author} {\bibfnamefont {J.~L.}\ \bibnamefont {Long}},
  \bibinfo {author} {\bibfnamefont {H.~S.}\ \bibnamefont {Ku}}, \bibinfo
  {author} {\bibfnamefont {R.~E.}\ \bibnamefont {Lake}}, \bibinfo {author}
  {\bibfnamefont {M.}~\bibnamefont {Bal}},\ and\ \bibinfo {author}
  {\bibfnamefont {D.~P.}\ \bibnamefont {Pappas}},\ }\bibfield  {title}
  {\bibinfo {title} {Overlap junctions for high coherence superconducting
  qubits},\ }\href {https://doi.org/10.1063/1.4993937} {\bibfield  {journal}
  {\bibinfo  {journal} {Applied Physics Letters}\ }\textbf {\bibinfo {volume}
  {111}},\ \bibinfo {pages} {032602} (\bibinfo {year} {2017})},\ \Eprint
  {https://arxiv.org/abs/https://pubs.aip.org/aip/apl/article-pdf/doi/10.1063/1.4993937/13614841/032602\_1\_online.pdf}
  {https://pubs.aip.org/aip/apl/article-pdf/doi/10.1063/1.4993937/13614841/032602\_1\_online.pdf}
  \BibitemShut {NoStop}%
\bibitem [{\citenamefont {Verjauw}\ \emph {et~al.}(2022)\citenamefont
  {Verjauw}, \citenamefont {Acharya}, \citenamefont {Van~Damme}, \citenamefont
  {Ivanov}, \citenamefont {Lozano}, \citenamefont {Mohiyaddin}, \citenamefont
  {Wan}, \citenamefont {Jussot}, \citenamefont {Vadiraj}, \citenamefont
  {Mongillo}, \citenamefont {Heyns}, \citenamefont {Radu}, \citenamefont
  {Govoreanu},\ and\ \citenamefont {Potočnik}}]{verjauw_path_2022}%
  \BibitemOpen
  \bibfield  {author} {\bibinfo {author} {\bibfnamefont {J.}~\bibnamefont
  {Verjauw}}, \bibinfo {author} {\bibfnamefont {R.}~\bibnamefont {Acharya}},
  \bibinfo {author} {\bibfnamefont {J.}~\bibnamefont {Van~Damme}}, \bibinfo
  {author} {\bibfnamefont {T.}~\bibnamefont {Ivanov}}, \bibinfo {author}
  {\bibfnamefont {D.~P.}\ \bibnamefont {Lozano}}, \bibinfo {author}
  {\bibfnamefont {F.~A.}\ \bibnamefont {Mohiyaddin}}, \bibinfo {author}
  {\bibfnamefont {D.}~\bibnamefont {Wan}}, \bibinfo {author} {\bibfnamefont
  {J.}~\bibnamefont {Jussot}}, \bibinfo {author} {\bibfnamefont {A.~M.}\
  \bibnamefont {Vadiraj}}, \bibinfo {author} {\bibfnamefont {M.}~\bibnamefont
  {Mongillo}}, \bibinfo {author} {\bibfnamefont {M.}~\bibnamefont {Heyns}},
  \bibinfo {author} {\bibfnamefont {I.}~\bibnamefont {Radu}}, \bibinfo {author}
  {\bibfnamefont {B.}~\bibnamefont {Govoreanu}},\ and\ \bibinfo {author}
  {\bibfnamefont {A.}~\bibnamefont {Potočnik}},\ }\bibfield  {title} {\bibinfo
  {title} {Path toward manufacturable superconducting qubits with relaxation
  times exceeding 0.1 ms},\ }\href {https://doi.org/10.1038/s41534-022-00600-9}
  {\bibfield  {journal} {\bibinfo  {journal} {npj Quantum Information}\
  }\textbf {\bibinfo {volume} {8}},\ \bibinfo {pages} {93} (\bibinfo {year}
  {2022})}\BibitemShut {NoStop}%
\bibitem [{\citenamefont {Ambegaokar}\ and\ \citenamefont
  {Baratoff}(1963)}]{ambegaokar_tunneling_1963}%
  \BibitemOpen
  \bibfield  {author} {\bibinfo {author} {\bibfnamefont {V.}~\bibnamefont
  {Ambegaokar}}\ and\ \bibinfo {author} {\bibfnamefont {A.}~\bibnamefont
  {Baratoff}},\ }\bibfield  {title} {\bibinfo {title} {Tunneling {Between}
  {Superconductors}},\ }\href {https://doi.org/10.1103/PhysRevLett.10.486}
  {\bibfield  {journal} {\bibinfo  {journal} {Physical Review Letters}\
  }\textbf {\bibinfo {volume} {10}},\ \bibinfo {pages} {486} (\bibinfo {year}
  {1963})}\BibitemShut {NoStop}%
\bibitem [{\citenamefont {Chiang}\ and\ \citenamefont
  {Wager}(2018)}]{Chiang_electronic_2018}%
  \BibitemOpen
  \bibfield  {author} {\bibinfo {author} {\bibfnamefont {T.-H.}\ \bibnamefont
  {Chiang}}\ and\ \bibinfo {author} {\bibfnamefont {J.~F.}\ \bibnamefont
  {Wager}},\ }\bibfield  {title} {\bibinfo {title} {Electronic conduction
  mechanisms in insulators},\ }\href {https://doi.org/10.1109/TED.2017.2776612}
  {\bibfield  {journal} {\bibinfo  {journal} {IEEE Transactions on Electron
  Devices}\ }\textbf {\bibinfo {volume} {65}},\ \bibinfo {pages} {223}
  (\bibinfo {year} {2018})}\BibitemShut {NoStop}%
\bibitem [{\citenamefont {Perkins}\ \emph {et~al.}(2018)\citenamefont
  {Perkins}, \citenamefont {Jenkins}, \citenamefont {Chiang}, \citenamefont
  {Mansergh}, \citenamefont {Gouliouk}, \citenamefont {Kenane}, \citenamefont
  {Wager}, \citenamefont {Conley},\ and\ \citenamefont
  {Keszler}}]{Perkins_demonstration_2018}%
  \BibitemOpen
  \bibfield  {author} {\bibinfo {author} {\bibfnamefont {C.~K.}\ \bibnamefont
  {Perkins}}, \bibinfo {author} {\bibfnamefont {M.~A.}\ \bibnamefont
  {Jenkins}}, \bibinfo {author} {\bibfnamefont {T.-H.}\ \bibnamefont {Chiang}},
  \bibinfo {author} {\bibfnamefont {R.~H.}\ \bibnamefont {Mansergh}}, \bibinfo
  {author} {\bibfnamefont {V.}~\bibnamefont {Gouliouk}}, \bibinfo {author}
  {\bibfnamefont {N.}~\bibnamefont {Kenane}}, \bibinfo {author} {\bibfnamefont
  {J.~F.}\ \bibnamefont {Wager}}, \bibinfo {author} {\bibfnamefont {J.~F.~J.}\
  \bibnamefont {Conley}},\ and\ \bibinfo {author} {\bibfnamefont {D.~A.}\
  \bibnamefont {Keszler}},\ }\bibfield  {title} {\bibinfo {title}
  {Demonstration of fowler–nordheim tunneling in simple solution-processed
  thin films},\ }\href {https://doi.org/10.1021/acsami.8b08986} {\bibfield
  {journal} {\bibinfo  {journal} {ACS Applied Materials \& Interfaces}\
  }\textbf {\bibinfo {volume} {10}},\ \bibinfo {pages} {36082} (\bibinfo {year}
  {2018})},\ \Eprint
  {https://arxiv.org/abs/https://doi.org/10.1021/acsami.8b08986}
  {https://doi.org/10.1021/acsami.8b08986} \BibitemShut {NoStop}%
\bibitem [{\citenamefont {Naghiloo}(2019)}]{naghiloo}%
  \BibitemOpen
  \bibfield  {author} {\bibinfo {author} {\bibfnamefont {M.}~\bibnamefont
  {Naghiloo}},\ }\href {https://arxiv.org/abs/1904.09291} {\bibinfo {title}
  {Introduction to experimental quantum measurement with superconducting
  qubits}} (\bibinfo {year} {2019}),\ \Eprint
  {https://arxiv.org/abs/1904.09291} {arXiv:1904.09291 [quant-ph]} \BibitemShut
  {NoStop}%
\bibitem [{\citenamefont {Krantz}\ \emph {et~al.}(2019)\citenamefont {Krantz},
  \citenamefont {Kjaergaard}, \citenamefont {Yan}, \citenamefont {Orlando},
  \citenamefont {Gustavsson},\ and\ \citenamefont {Oliver}}]{Krantz}%
  \BibitemOpen
  \bibfield  {author} {\bibinfo {author} {\bibfnamefont {P.}~\bibnamefont
  {Krantz}}, \bibinfo {author} {\bibfnamefont {M.}~\bibnamefont {Kjaergaard}},
  \bibinfo {author} {\bibfnamefont {F.}~\bibnamefont {Yan}}, \bibinfo {author}
  {\bibfnamefont {T.~P.}\ \bibnamefont {Orlando}}, \bibinfo {author}
  {\bibfnamefont {S.}~\bibnamefont {Gustavsson}},\ and\ \bibinfo {author}
  {\bibfnamefont {W.~D.}\ \bibnamefont {Oliver}},\ }\bibfield  {title}
  {\bibinfo {title} {A quantum engineer's guide to superconducting qubits},\
  }\href {https://doi.org/10.1063/1.5089550} {\bibfield  {journal} {\bibinfo
  {journal} {Applied Physics Reviews}\ }\textbf {\bibinfo {volume} {6}},\
  \bibinfo {pages} {021318} (\bibinfo {year} {2019})},\ \Eprint
  {https://arxiv.org/abs/https://pubs.aip.org/aip/apr/article-pdf/doi/10.1063/1.5089550/16667201/021318\_1\_online.pdf}
  {https://pubs.aip.org/aip/apr/article-pdf/doi/10.1063/1.5089550/16667201/021318\_1\_online.pdf}
  \BibitemShut {NoStop}%
\bibitem [{\citenamefont {Niu}\ \emph {et~al.}(2023)\citenamefont {Niu},
  \citenamefont {Zhang}, \citenamefont {Liu}, \citenamefont {Qiu},
  \citenamefont {Huang}, \citenamefont {Huang}, \citenamefont {Jia},
  \citenamefont {Liu}, \citenamefont {Tao}, \citenamefont {Wei}, \citenamefont
  {Zhou}, \citenamefont {Zou}, \citenamefont {Chen}, \citenamefont {Deng},
  \citenamefont {Deng}, \citenamefont {Hu}, \citenamefont {Hu}, \citenamefont
  {Li}, \citenamefont {Tan}, \citenamefont {Xu}, \citenamefont {Yan},
  \citenamefont {Yan}, \citenamefont {Liu}, \citenamefont {Zhong},
  \citenamefont {Cleland},\ and\ \citenamefont {Yu}}]{Niu_low_2023}%
  \BibitemOpen
  \bibfield  {author} {\bibinfo {author} {\bibfnamefont {J.}~\bibnamefont
  {Niu}}, \bibinfo {author} {\bibfnamefont {L.}~\bibnamefont {Zhang}}, \bibinfo
  {author} {\bibfnamefont {Y.}~\bibnamefont {Liu}}, \bibinfo {author}
  {\bibfnamefont {J.}~\bibnamefont {Qiu}}, \bibinfo {author} {\bibfnamefont
  {W.}~\bibnamefont {Huang}}, \bibinfo {author} {\bibfnamefont
  {J.}~\bibnamefont {Huang}}, \bibinfo {author} {\bibfnamefont
  {H.}~\bibnamefont {Jia}}, \bibinfo {author} {\bibfnamefont {J.}~\bibnamefont
  {Liu}}, \bibinfo {author} {\bibfnamefont {Z.}~\bibnamefont {Tao}}, \bibinfo
  {author} {\bibfnamefont {W.}~\bibnamefont {Wei}}, \bibinfo {author}
  {\bibfnamefont {Y.}~\bibnamefont {Zhou}}, \bibinfo {author} {\bibfnamefont
  {W.}~\bibnamefont {Zou}}, \bibinfo {author} {\bibfnamefont {Y.}~\bibnamefont
  {Chen}}, \bibinfo {author} {\bibfnamefont {X.}~\bibnamefont {Deng}}, \bibinfo
  {author} {\bibfnamefont {X.}~\bibnamefont {Deng}}, \bibinfo {author}
  {\bibfnamefont {C.}~\bibnamefont {Hu}}, \bibinfo {author} {\bibfnamefont
  {L.}~\bibnamefont {Hu}}, \bibinfo {author} {\bibfnamefont {J.}~\bibnamefont
  {Li}}, \bibinfo {author} {\bibfnamefont {D.}~\bibnamefont {Tan}}, \bibinfo
  {author} {\bibfnamefont {Y.}~\bibnamefont {Xu}}, \bibinfo {author}
  {\bibfnamefont {F.}~\bibnamefont {Yan}}, \bibinfo {author} {\bibfnamefont
  {T.}~\bibnamefont {Yan}}, \bibinfo {author} {\bibfnamefont {S.}~\bibnamefont
  {Liu}}, \bibinfo {author} {\bibfnamefont {Y.}~\bibnamefont {Zhong}}, \bibinfo
  {author} {\bibfnamefont {A.~N.}\ \bibnamefont {Cleland}},\ and\ \bibinfo
  {author} {\bibfnamefont {D.}~\bibnamefont {Yu}},\ }\bibfield  {title}
  {\bibinfo {title} {Low-loss interconnects for modular superconducting quantum
  processors},\ }\href {https://doi.org/10.1038/s41928-023-00925-z} {\bibfield
  {journal} {\bibinfo  {journal} {Nature Electronics}\ }\textbf {\bibinfo
  {volume} {6}},\ \bibinfo {pages} {235–241} (\bibinfo {year}
  {2023})}\BibitemShut {NoStop}%
\bibitem [{\citenamefont {Smith}\ \emph {et~al.}(2022)\citenamefont {Smith},
  \citenamefont {Ravi}, \citenamefont {Baker},\ and\ \citenamefont
  {Chong}}]{smith_scaling_2022}%
  \BibitemOpen
  \bibfield  {author} {\bibinfo {author} {\bibfnamefont {K.~N.}\ \bibnamefont
  {Smith}}, \bibinfo {author} {\bibfnamefont {G.~S.}\ \bibnamefont {Ravi}},
  \bibinfo {author} {\bibfnamefont {J.~M.}\ \bibnamefont {Baker}},\ and\
  \bibinfo {author} {\bibfnamefont {F.~T.}\ \bibnamefont {Chong}},\ }\bibfield
  {title} {\bibinfo {title} {Scaling superconducting quantum computers with
  chiplet architectures},\ }in\ \href
  {https://doi.org/10.1109/MICRO56248.2022.00078} {\emph {\bibinfo {booktitle}
  {2022 55th IEEE/ACM International Symposium on Microarchitecture (MICRO)}}}\
  (\bibinfo {year} {2022})\ pp.\ \bibinfo {pages} {1092--1109}\BibitemShut
  {NoStop}%
\bibitem [{\citenamefont {Field}\ \emph {et~al.}(2024)\citenamefont {Field},
  \citenamefont {Chen}, \citenamefont {Scharmann}, \citenamefont {Sete},
  \citenamefont {Oruc}, \citenamefont {Vu}, \citenamefont {Kosenko},
  \citenamefont {Mutus}, \citenamefont {Poletto},\ and\ \citenamefont
  {Bestwick}}]{Field_Modular_2024}%
  \BibitemOpen
  \bibfield  {author} {\bibinfo {author} {\bibfnamefont {M.}~\bibnamefont
  {Field}}, \bibinfo {author} {\bibfnamefont {A.~Q.}\ \bibnamefont {Chen}},
  \bibinfo {author} {\bibfnamefont {B.}~\bibnamefont {Scharmann}}, \bibinfo
  {author} {\bibfnamefont {E.~A.}\ \bibnamefont {Sete}}, \bibinfo {author}
  {\bibfnamefont {F.}~\bibnamefont {Oruc}}, \bibinfo {author} {\bibfnamefont
  {K.}~\bibnamefont {Vu}}, \bibinfo {author} {\bibfnamefont {V.}~\bibnamefont
  {Kosenko}}, \bibinfo {author} {\bibfnamefont {J.~Y.}\ \bibnamefont {Mutus}},
  \bibinfo {author} {\bibfnamefont {S.}~\bibnamefont {Poletto}},\ and\ \bibinfo
  {author} {\bibfnamefont {A.}~\bibnamefont {Bestwick}},\ }\bibfield  {title}
  {\bibinfo {title} {Modular superconducting-qubit architecture with a
  multichip tunable coupler},\ }\href
  {https://doi.org/10.1103/PhysRevApplied.21.054063} {\bibfield  {journal}
  {\bibinfo  {journal} {Phys. Rev. Appl.}\ }\textbf {\bibinfo {volume} {21}},\
  \bibinfo {pages} {054063} (\bibinfo {year} {2024})}\BibitemShut {NoStop}%
\end{thebibliography}%
\end{document}